\documentclass[fleqn,usenatbib]{mnras}

\usepackage[T1]{fontenc}

\usepackage{blindtext}
\usepackage{graphicx}	
\usepackage{amsmath}	
\usepackage{amssymb}	
\usepackage{pifont}
\usepackage{pdflscape}
\usepackage{rotating}
\usepackage{newtxtext,newtxmath}

\title[An ALMA view of 11 DSFGs at $z\sim2$]{An ALMA view of 11 Dusty Star Forming Galaxies at the peak of Cosmic Star Formation History} 

\author[L. Pantoni et al.]{L. Pantoni,$^{1,2,5}$\thanks{E-mail: lpantoni@sissa.it}
M. Massardi,$^{5}$
A. Lapi,$^{1,2,3,5}$
D. Donevski,$^{1,3}$
Q. D'Amato$,^{5}$
M. Giulietti,$^{1,5}$
F. Pozzi,$^{6,7}$
\newauthor{M. Talia,$^{6,7}$
C. Vignali,$^{6,7}$
A. Cimatti,$^{6,8}$
L. Silva,$^{4}$
A. Bressan,$^{1}$
and T. Ronconi$^{1,2,3}$}
\\
$^{1}$SISSA - ISAS, Via Bonomea 265,  Trieste 34136, Italy\\
$^{2}$INFN - Sezione di Trieste, via Valerio 2, Trieste 34127, Italy\\
$^{3}$IFPU - Institute for fundamental physics of the Universe, Via Beirut 2, 34014 Trieste, Italy\\
$^{4}$INAF - Osservatorio Astronomico di Trieste, Via Giambattista Tiepolo, 11, 34131 Trieste, Italy\\
$^{5}$INAF - Istituto di Radioastronomia - Italian ARC, Via Piero Gobetti 101, I-40129 Bologna, Italy\\
$^{6}$DIFA - Dipartimento di Fisica e Astronomia, Universit\`{a} degli Studi di Bologna, Via Berti Pichat 6/2, I-40127 Bologna, Italy\\
$^{7}$INAF - Osservatorio di Astrofisica e Scienza dello Spazio di Bologna, Via Gobetti 93/3, I-40129 Bologna, Italy\\
$^{8}$INAF - Osservatorio Astrofisico di Arcetri, Largo E. Fermi, 50125, Firenze, Italy
}

\date{Accepted 2021 August 11. Received 2021 August 9; in original form 2021 June 15.}

\pubyear{2021}
\newcommand{\xmark}{\ding{55}}
\newcommand{\cmark}{\ding{51}}
\begin{document}
\label{firstpage}
\pagerange{\pageref{firstpage}--\pageref{lastpage}}
\maketitle

\begin{abstract}
We present the ALMA view of 11 main-sequence DSFGs, (sub-)millimeter selected in the GOODS-S field, and spectroscopically confirmed to be at the peak of Cosmic SFH ($z\sim2$). Our study combines the analysis of galaxy SED with ALMA continuum and CO spectral emission, by using ALMA Science Archive products at the highest spatial resolution currently available for our sample ($\Delta\theta\lesssim 1$ arcsec). We include galaxy multi-band images and photometry (in the optical, radio and X-rays) to investigate the interlink between dusty, gaseous and stellar components and the eventual presence of AGN. We use multi-band sizes and morphologies to gain an insight on the processes that lead galaxy evolution, e.g. gas condensation, star formation, AGN feedback. The 11 DSFGs are very compact in the (sub-)millimeter (median r$_{\rm ALMA}=1.15$ kpc), while the optical emission extends to larger radii (median r$_{\rm H}/$r$_{\rm ALMA}=2.05$). CO lines reveal the presence of a rotating disc of molecular gas, but we can not exclude either the presence of interactions and/or molecular outflows. Images at higher (spectral and spatial) resolution are needed to disentangle from the possible scenarios. Most of the galaxies are caught in the \textit{compaction} phase, when gas cools and falls into galaxy centre, fuelling the dusty burst of star formation and the growing nucleus. We expect these DSFGs to be the high-z star-forming counterparts of massive quiescent galaxies. Some features of CO emission in three galaxies are suggestive of forthcoming/ongoing AGN feedback, that is thought to trigger the morphological transition from star-forming disks to ETGs.
\end{abstract}




\section{Introduction}\label{sec:intro}

Dusty Star-Forming Galaxies (DSFGs) at high-redshift ($z>1$) have been recognised as a crucial population to characterize massive galaxy evolution across the Universe and constrain the Cosmic Star Formation History (SFH) and stellar mass assembly out to redshift $> 3$ \citep[e.g.,][]{MadauDickinson2014:2014ARA&A..52..415M, CaseyNarayananCooray2014:2014PhR...541...45C}. In the last decade, many steps forwards have been taken to obtain a more comprehensive picture of their nature, thanks to the advent of a new generation of (sub-)millimeter/radio telescopes, with increased resolution and sensitivity, e.g. ALMA and JVLA, and to numerous multi-wavelength surveys that are essential to characterize DSFG broad-band emission. For example, the Great Observatories Origins Survey South \citep[GOODS-S,][]{Dickinson2001:2001AAS...198.2501D,Giavalisco2004:2004ApJ...600L..93G} field - one of the most studied - was sampled in the X-ray with \textit{Chandra} \citep[][]{Xue2011:2011ApJS..195...10X, Luo2017:2017ApJS..228....2L} and XMM-Newton \citep[][]{Comastri2011:2011A&A...526L...9C}; in the optical/near-infrared with HST \citep[i.e., the HUDF survey,][]{Beckwith2006:2006AJ....132.1729B} and VLT MUSE \citep[][]{Bacon2017:2017A&A...608A...1B}; in the infrared with \textit{Herschel} \citep[PEP \& HerMES,][]{Lutz2011:2011A&A...532A..90L, Oliver2012:2012MNRAS.424.1614O} and \textit{Spitzer} \citep[][]{Labbe2015:2015ApJS..221...23L}; in the (sub-)millimeter with LABOCA \citep[LESS,][]{Weiss2009:2009ApJ...707.1201W}, AzTEC/ASTE \citep[][]{Scott2010:2010MNRAS.405.2260S} and ALMA \citep[][]{Aravena2016a:2016MNRAS.457.4406A,Aravena2016b:2016ApJ...833...68A,Walter2016:2016ApJ...833...67W, Dunlop2017:2017MNRAS.466..861D}; in the radio band with VLA \citep[][]{Kellermann2008:2008ApJS..179...71K, Miller2013:2013ApJS..205...13M, Rujopakarn2016:2016ApJ...833...12R,Fujimoto2017:2017ApJ...850...83F}. 
However, we are far away from fully physically characterizing the high-z DSFG population. In particular, we lack of a self-consistent explanation of both the mechanisms triggering their intense burst of dust-obscured star formation, with Star Formation Rates (SFRs) $\gtrsim 100$ M$_\odot$ yr$^{-1}$, and the processes driving their subsequent evolution. In order to reach this goal it is essential, on the one hand, to characterize their integral physical properties (i.e., stellar mass, dust and gas content, gas metallicity, the activity of the central Super Massive Black Hole, SMBH) by sampling their Spectral Energy Distributions \citep[SEDs; see e.g.,][]{Bethermin2014:2014A&A...567A.103B, Magdis2012;2012ApJ...760....6M, Malek2018:2018A&A...620A..50M, Bianchini2019:2019ApJ...871..136B, Donevski2020:2020A&A...644A.144D, Dudzeviciute2020:2020MNRAS.494.3828D}. 
On the other hand, spatially-resolved information provides a detailed and precise description of the main baryonic processes occurring inside these galaxies and can be useful to determine their respective importance in driving DSFG evolution. In particular the latter objective can be reached by exploiting high resolution imaging of objects that do not appear peculiar in their overall behaviour, so that they could represent the bulk of the $z\sim2$ DSFG population. The detection of molecular spectral lines and the study of galaxy multi-band size and morphology would allow us to investigate galaxy environment and the gas phase properties, such as its kinematics \citep[e.g.,][]{Tadaki2015:2015ApJ...811L...3T,Talia2018:2018MNRAS.476.3956T, Chen2017:2017ApJ...846..108C,  Hodge2019:2019ApJ...876..130H}. This approach is fundamental to asses the presence of an AGN, characterize its impact on the host galaxy \citep[i.e., AGN \textit{feeding \& feedback} cycle; e.g.,][]{Bischetti2021:2021A&A...645A..33B} and, on a statistical basis, on DSFG evolution.

This work is meant to complement the analysis presented in a previous paper from our group \citep[][]{Pantoni2021:2021MNRAS.504..928P} on 11 (sub-)millimeter selected DSFGs at $z_{\rm spec}\sim2$ by focusing on the ALMA view of the galaxies. Our 11 DSFGs are located in the GOODS-S field and have an almost complete sampling of their multi-wavelength broad-band emission, from the X-rays to the radio band, that was extensively studied in \citet{Pantoni2021:2021MNRAS.504..928P}. In a nutshell, we modelled the galaxy optical-to-millimeter SED by performing a multi-component fitting to the available multi-wavelength photometry with the Code Investigating GALaxy Emission \citep[CIGALE;][]{Boquien2019:2019A&A...622A.103B}. We combined \cite{BruzualCharlot2003:2003MNRAS.344.1000B} stellar libraries, \cite{LoFaro2017:2017MNRAS.472.1372L} double power-law (describing stellar attenuation due to dust extinction) and the mid- and far-IR dust emission model by \cite{Casey2012:2012MNRAS.425.3094C} with new physical motivated prescriptions for dust absorption in the birth molecular clouds of stars. Then, we included in our analysis other information coming from integrated CO spectroscopy and galaxy emission in the X-rays and radio band, providing a self-consistent picture of the ongoing processes concerning galaxy baryonic components, i.e., stars, inter-stellar medium (ISM) - and its molecular and dusty phases - and central SMBH. To this aim we referred to one possible scenario for galaxy formation and evolution, the so-called \textit{in-situ} galaxy-BH co-evolution scenario \citep[see e.g.][]{Shi2017:2017ApJ...843..105S, Mancuso2016a:2016ApJ...823..128M, Mancuso2016b:2016ApJ...833..152M, Lapi2018a:2018ApJ...857...22L, Pantoni2019:2019ApJ...880..129P}.

In the following we present the spatially-resolved analysis of millimeter and (sub-)millimeter continuum and CO emission lines, as observed with ALMA, in order to derive the size of interstellar dust and CO emission, the molecular gas content and its kinematics. We complement the ALMA view of the galaxies with resolved multi-band information (size and morphology) found in the literature \citep[e.g.][]{vanderWel2012:2012ApJS..203...24V, Targett2013:2013MNRAS.432.2012T, Rujopakarn2016:2016ApJ...833...12R, Rujopakarn2019:2019ApJ...882..107R, Kaasinen2020:2020ApJ...899...37K}. Comparing galaxy morphology and size in different spectral bands is informative of the processes that are driving the ongoing burst of star formation, the accretion of central SMBH and its eventual activity and can tell us if and how the AGN affects the whole host galaxy evolution. The optical rest-frame light samples the stellar component and thus its spatial distribution into galaxy, rest-frame MIR and FIR emission samples the interstellar dust, while the CO emission traces the molecular gas. The relative size of these components provides a hint on their origin, their role in fuelling the star formation and the central SMBH accretion. The multi-band morphology (e.g., isolated or disturbed) is a result of gas condensation and star formation processes occurring inside the galaxies and of eventual interactions with the surrounding ambient and/or galaxy companions  \citep[][]{Lacey2016:2016MNRAS.462.3854L, Calura2017:2017MNRAS.465...54C, Popping2017:2017MNRAS.471.3152P, Lapi2018a:2018ApJ...857...22L, Dave2019:2019MNRAS.486.2827D}.

By including in our picture the outcomes of the SED analysis presented in \citet{Pantoni2021:2021MNRAS.504..928P} we provide a consistent and physically-motivated interpretation of the observations in the broad context of galaxy formation and evolution, that encompasses photometry, spectroscopy and imaging at high-resolution ($\Delta\theta\lesssim$ 1 arcsec).

This article is organised as follows. 
In Section \ref{sec:thesample} we summarize the main selection criteria used to build our sample of 11 DSFGs at $z\sim2$ including a brief characterization of the galaxies based on previous results \citep[][]{Pantoni2021:2021MNRAS.504..928P}; in Section \ref{sec:continuumemission} and Section \ref{sec:spectroscopicalemission} we analyse the ALMA continuum and spectroscopic emission of the sources, derive dust and molecular gas size and spatial distribution and the total content of molecular gas, and provide a possible description of the gas kinematics from CO lines, when detected. In Section \ref{sec:discussion} we discuss the main results for the whole sample in the broad context of galaxy formation and evolution. In Section \ref{sec:conclusions} we summarize the outcomes and outline our conclusions. In Appendix \ref{appendix} we include a panchromatic analysis of the individual galaxies evolution.

Throughout this work, we adopt the standard flat $\Lambda$CDM cosmology \cite{Planck2018:2020A&A...641A...6P} with rounded parameter values: matter density $\Omega_{\rm M} = 0.32$, dark energy density $\Omega_{\rm \Lambda} = 0.63$, baryon density $\Omega_{\rm b} = 0.05$, Hubble constant $H_0$ = 100 $h$ km s$^{-1}$ Mpc$^{-1}$ with $h$ = 0.67, and mass variance $\sigma_8$ = 0.81 on a scale of 8 $h^{-1}$ Mpc.

\section{The sample}\label{sec:thesample}

\begin{table*}
 \centering
 \caption{Main galaxy properties from literature. In the order: (sub-)millimeter source ID; IAU ID; spectroscopic redshift (z); presence of an AGN by \citet[][]{Luo2017:2017ApJS..228....2L}; H$_{160}$ circularized size (r$_{\rm H}$) derived from \citet[][]{Rujopakarn2019:2019ApJ...882..107R} for UDF1  \citep[assuming the axes ratio by][]{vanderWel2012:2012ApJS..203...24V}, and 
 computed by \citet{Pantoni2021:2021MNRAS.504..928P} from the
 effective S\'{e}rsic half-light semi-axis by \citet[][]{vanderWel2012:2012ApJS..203...24V} for all the other sources}; stellar mass (M$_\star$), star formation rate (SFR) and dust mass (M$_{\rm dust}$) by \citet{Pantoni2021:2021MNRAS.504..928P}.\label{tab:galaxy_properties}
 \resizebox{2.1\columnwidth}{!}{
\hspace{-0.5cm}
  \begin{tabular}{llcccccc}
  \hline
   \textbf{ID} & \textbf{IAU ID} & \textbf{z} & \textbf{AGN} & $\mathbf{r_H}$ & $\mathbf{M_\star}$ & \textbf{SFR} & $\mathbf{M_{dust}}$\\
    &  &  &  & [kpc] & [$10^{10}$ M$_\odot$] & [M$_\odot$ yr$^{-1}$] & [$10^{8}$ M$_\odot$] \\
   \hline
    UDF1 & J033244.01-274635.2 & $2.698\pm0.001^{(a)}$ & \cmark & 2.6 & $8\pm1$ & $352\pm18$ & $5.6\pm0.2$ \\
    UDF3 & J033238.53-274634.6 & ${2.544\pm^{+0.001}_{-0.002}}^{(a)}$ & \cmark & $1.6$ & $9\pm1$ & $519\pm38$ & $4.0\pm1.6$ \\
    UDF5 & J033236.94-274726.8 & $1.759\pm0.008^{(b)}$ & \textbf{nd} & $2.3$ & $2.4\pm0.3$ & $85\pm6$ & $4.6\pm2.2$ \\
    UDF8 & J033239.74-274611.4 & ${1.5510^{+0.0014}_{-0.0005}}^{(a)}$ & \cmark & $5.7$ & $6.5\pm0.3$ & $100\pm5$ & $2.4\pm1.3$ \\
    UDF10 & J033240.73-274749.4 & $2.086\pm0.006^{(b)}$ & \xmark & $2.0$ & $2.5\pm0.3$ & $41\pm5$ & $1.9\pm1.3$ \\
    UDF11 & J033240.06-274755.5 & $1.9962\pm0.0014^{(c,\,d)}$ & \xmark & $4.5$ & $6.4\pm0.9$ & $241\pm19$ & $1.46\pm0.66$ \\
    UDF13 & J033235.07-274647.6 & $2.497\pm0.008^{(b)}$ & \cmark & $1.2$ & $6.5\pm1.4$ & $111\pm17$ & $1.20\pm0.68$ \\
    ALESS067.1 & J033243.19-275514.3 & ${2.1212^{+0.0014}_{-0.0005}}^{(a)}$ & \cmark & $6.5$ & $29\pm3$ & $485\pm24$ & $10\pm4$ \\
    AzTEC.GS25 & J033246.83-275120.9 & $2.292\pm0.001^{(e)}$ & \cmark & $1.8$ & $8\pm2$ & $401\pm20$ & $13.6\pm8.2$ \\
    AzTEC.GS21 & J033247.59-274452.3 & $1.910\pm0.001^{(f)}$ & \cmark & $3.7$ & $18\pm2$ & $360\pm18$ & $5.8\pm1.4$ \\
    AzTEC.GS22 & J033212.55-274306.1 & $1.794\pm0.005^{(g)}$ & \textbf{nd} & $3.2$ & $5.7\pm0.5$ & $91\pm5$ & $14\pm8$ \\
  \hline
  \end{tabular}
  }
References for spectroscopic redshifts:
$a$) This work; 
$b$) \citet{Momcheva2016:2016ApJS..225...27M}; 
$c$) \citet{Kurk2013:2013A&A...549A..63K}; 
$d$) \citet{Dunlop2017:2017MNRAS.466..861D}; 
$e$) \citet{Popesso2009:2009A&A...494..443P};
$f$) \citet{Vanzella2008:2008A&A...478...83V}; 
$g$) \citet{Targett2013:2013MNRAS.432.2012T} and reference therein. 
\textbf{AGN} \citep[object classification by][]{Luo2017:2017ApJS..228....2L}: \cmark $\,$ means that the galaxy has a signature of AGN; \xmark $\,$ means that the object is classified as a normal star-forming galaxy; \textbf{nd} stands for \textit{not detected}.
\end{table*}

The sample is constituted by 11 DSFGs observed in the GOODS-S \citep[][]{Dickinson2001:2001AAS...198.2501D, Giavalisco2004:2004ApJ...600L..93G} field and spectroscopically-confirmed to be at the peak of Cosmic SFH (z $\sim2$). They were selected in the (sub-)millimeter regime requiring the following criteria to be fulfilled for each galaxy: 3 or more detections in the optical domain ($\lambda_{\rm obs}=0.3-1$ $\mu$m); 6 or more detections in the NIR+MIR bands ($\lambda_{\rm obs}=1-25$ $\mu$m); 2 or more detections in the FIR band ($\lambda_{\rm obs}=25-400$ $\mu$m);
spectroscopically confirmed redshift in the range $1.5<z<3$; one or more detections and/or upper limits in the radio and X-ray regimes. For the detailed description of source selection and multi-wavelength counterparts association we refer to \citet[][]{Pantoni2021:2021MNRAS.504..928P}. 

Thanks to the aforementioned selection criteria, in \citet[]{Pantoni2021:2021MNRAS.504..928P} we achieved a complete and accurate sampling of galaxy multi-wavelength broad-band emission (from X-rays to radio band), that was essential to derive galaxy integral properties (such as galaxy age, stellar mass, dust and gas content, the presence of an accreting/active central BH) by fitting their SED with CIGALE \citep[]{Boquien2019:2019A&A...622A.103B}, assuming a \cite{Chabrier2003:2003PASP..115..763C} Initial Mass Function (IMF).

Source IDs are listed in Tab. \ref{tab:galaxy_properties}, along with their spectroscopic redshift, their optical size \citep[i.e., circularized radius r$_{\rm H}$, derived by using the semi-major axis and axes ratio by][]{vanderWel2012:2012ApJS..203...24V, Rujopakarn2019:2019ApJ...882..107R}, the presence of an AGN \citep[by][]{Luo2017:2017ApJS..228....2L}, and some of their main astrophysical properties, i.e. stellar mass M$_\star$, SFR, interstellar dust mass M$_{\rm dust}$, already presented in \citet[][]{Pantoni2021:2021MNRAS.504..928P}. The 11 DSFGs are (almost) main-sequence objects \citep[i.e. staying within its $2\sigma$ scatter $\sim 0.4$ dex; see][]{Speagle2014:2014ApJS..214...15S}, with median M$_\star=6.5\times10^{10}$ M$_\odot$ (first quartile $5.7\times10^{10}$ M$_\odot$ and third quartile $9\times10^{10}$ M$_\odot$) and SFR $\sim250$ M$_\odot$ yr$^{-1}$ (first quartile 90 M$_\odot$ yr$^{-1}$ and third quartile 400 M$_\odot$ yr$^{-1}$). They are experiencing an intense and dusty burst of star formation (median L$_{\rm IR}\sim2\times10^{12}$ L$_\odot$), with a median duration $\tau_{\star}\sim750$ Myr (first quartile 300 Myr and third quartile 900 Myr). Despite their young age, the interstellar dust content is high (M$_{\rm dust}>10^8$ M$_\odot$) and possibly due to a very rapid enrichment of the ISM (on typical timescales of $10^7-10^8$ yr). The gas mass (i.e. median M$_{\rm gas}\sim4\times10^{10}$ M$_\odot$, derived from dust continuum), that fuels the dusty star formation, will be rapidly depleted over a median timescale $\tau_{\rm depl}\sim200$ Myr. Nine objects out of eleven have an X-ray luminous (L$_{2-10 {\rm keV}}\gtrsim10^{42}$ erg s$^{-1}$) counterpart in the \textit{Chandra} $\simeq7$ Ms catalog by \citet[][]{Luo2017:2017ApJS..228....2L} and two of them (UDF1 and UDF8) are clearly dominated in the X-rays by the active nucleus emission (L$_{2-10 {\rm keV}}\gtrsim10^{43}-10^{44}$ erg s$^{-1}$). Radio luminosities are consistent with the emission coming from galaxy star formation, suggesting that the AGN should be radio silent or quiet. 

\section{Continuum emission}\label{sec:continuumemission}

\begin{table*}
 \centering
 \caption{ALMA continuum observation setting and results. In the order: ID from literature; project code; member observing unit set ID; frequency band ($\nu_{\rm band}$); angular resolution (i.e., restoring beam, $\Delta \theta$); flux density (S$_{\nu}$); FWHM of the Gaussian fit of the angular size deconvolved from beam ($\theta_\nu$); linear circularized size (r$_{\rm ALMA}$).}
 \label{tab:continuum}
  \resizebox{2.1\columnwidth}{!}{
\hspace{-0.5cm}
  \begin{tabular}{lllccccc}
  \hline
  \textbf{ID}  & \textbf{Project code} & \textbf{Member ous} & $\mathbf{\nu_{band}}$ & $\mathbf{\Delta \theta=a_{\Delta\theta}\times b_{\Delta\theta}}$ & $\mathbf{S_\nu}$ & $\mathbf{\theta_\nu= a_{\nu}\times b_{\nu}}$ & $\mathbf{r_{ALMA}}$\\
   &  &  & [GHz] & [arcsec $\times$ arcsec] & [$\mu$Jy] & [arcsec $\times$ arcsec] & [kpc] \\
  \hline
   UDF1 & 2017.1.00001.S & A001/X1288/X4c3 & B7 ($335.50-351.48$) & $0.10\times0.09$ & $2900\pm300$& $0.123\pm0.006 \times 0.103\pm0.005$ & 0.46 \\
   UDF3 & 2017.1.00001.S & A001/X1288/X4c7 & B7 ($335.50-351.48$) & $0.08\times0.07$ & $1600\pm200$ & $0.203\pm0.014\times0.111\pm0.009$ & 0.62\\
   UDF5 & 2012.1.00173.S & A002/X5a9a13/X7e0 & B6 ($211.21-231.20$) & $0.62\times0.52$ & $311\pm49^D$ & $0.62\times0.52^{sb}$  & $<2.5$ \\
   UDF8 & 2012.1.00173.S & A002/X5a9a13/X7e0 & B6 ($211.21-231.20$) & $0.62\times0.52$ & $208\pm46^D$ & $1.42\pm0.35\times0.66\pm0.19$ & 4.1 \\ 
   UDF10 & 2012.1.00173.S & A002/X5a9a13/X7e0 & B6 ($211.21-231.20$) & $0.62\times0.52$ & $184\pm46^D$ & $0.62\times0.52^{sb}$  & $<2.5$ \\
   UDF11 & 2012.1.00173.S & A002/X5a9a13/X7e0 & B6 ($211.21-231.20$) & $0.62\times0.52$ & $186\pm46^D$ & $1.02\pm0.28\times0.61\pm0.21$& 3.4 \\ 
   UDF13 & 2015.1.01074.S$^A$ & A001/X2d8/Xfd & B7 ($335.50-351.48$) & $0.17\times0.15$ & $910\pm170$ & $0.17\times0.15^{sb}$ & $< 0.65$\\
   ALESS067.1 & 2012.1.00307.S & A002/X7d1738/X103 & B7 ($336.00-351.98$) & $0.14\times0.12$ & $4500\pm 400$ & $0.35\pm0.05 \times 0.18\pm0.03$ & 1.1 \\ 
   AzTEC.GS25 & 2012.1.00983.S & A002/X7d1738/X169
 & B7 ($336.00-351.98$) & $0.20\times0.16$ &  $5900\pm500$ & $0.38\pm0.03 \times 0.22\pm0.02$ & 1.2 \\ 
   AzTEC.GS21 & 2015.1.00098.S$^A$ & A001/X2fe/Xaea & B6 ($244.13-262.99$) & $0.18\times0.16$  & $954\pm74$ & $0.18\times0.16^{sb}$  & $< 0.7$ \\
   AzTEC.GS22 & 2017.1.01347.S & A001/X12a3/X80e & B9 ($662.53-685.56$) & $0.49\times 0.33$ & $6400\pm880$ &  $0.49\times 0.33^{sb}$& $<1.7$\\ 
  \hline
  \end{tabular}
  }
  For unresolved sources labelled with $^{sb}$ the synthesized beam size is reported and treated as an \textit{upper limits} to the actual source size. ALMA observations labelled with the apex $A$ have been re-imaged in the context of ARI-L project \citep[https://almascience.eso.org/alma-data/aril; see also][]{Massardi2021:2021arXiv210711071M}. Flux densities labelled with the apex $D$ are taken from \citet{Dunlop2017:2017MNRAS.466..861D}.
\end{table*}

\begin{table*}
 \centering
 \caption{ALMA observed CO emission. In the order: ID from literature; project code; member ous ID; velocity resolution ($\Delta v$); angular resolution (i.e., restoring beam, $\Delta \theta$); upper level of the CO transition (J$^{\rm up}$); FWHM; CO line intensity (I$_{\rm CO}$); CO angular size ($\theta_\nu$); CO linear circularized size (r$_{\rm CO}$). Quantities preceded with a "$<$" are \textit{upper limits}.
 }\label{tab:COemission}
 \resizebox{2.1\columnwidth}{!}{
\hspace{-0.5cm}
  \begin{tabular}{lllcccccccc}
  \hline
  \textbf{ID}  & \textbf{Project code} & \textbf{Member ous} & $\mathbf{\Delta v}$ & $\mathbf{\Delta \theta = a_\theta\times b_\theta}$  & $\mathbf{J^{up}}$ & $\mathbf{\nu_{obs}}$ & $\mathbf{FWHM}$ & $\mathbf{I_{CO}}$  & $\mathbf{\theta_\nu}$ & $\mathbf{r_{CO}}$ \\
   &  &  & [km/s] & [arcsec $\times$ arcsec] &  & [GHz] & [km/s] & [mJy km s$^{-1}$] & [arcsec] & [kpc] \\
  \hline
   UDF1 & 2017.1.00270.S & A001/X1288/X484 & 0.85 & $0.662\times0.440$ & 3 & 93.5 & 170 & $79.9\pm7.4$ & $<0.812$ & $<3.3$ \\ 
   UDF3	& 2016.1.00324.L & A001/X87c/X20a & 59.4 & $1.34\times1.171$ & 3 & 97.6 & 190; 390 & $492\pm28$ & 0.326 & 1.35 \\
   UDF8	& 2016.1.00324.L & A001/X87c/X20e & 64.1 & $2.002\times1.648$ & 2 & 90.4 & 148; 379 & $378\pm25$ & 0.800 & 3.45  \\ 
   ALESS067.1 & 2016.1.00564.S  & A001/X879/Xd9 &  21.2 & $2.004\times1.286$ & 3 & 110.8 & 115; 314 & $774\pm120$ & 0.842 & 3.6 \\ 
   & 2019.2.00246.S & A001/X14c3/Xaf4 & 21.0 & $8.330\times4.363$ & 6 & 221.6 & 45 & $620\pm70$ & $<3$ & $<26$  \\
  \hline
  \end{tabular}
  }
\end{table*}

Continuum (sub-)millimeter light of $z\sim2$ star forming galaxies traces the thermal emission coming from interstellar dust grains, that are heated by newly formed stars \citep[e.g.,][]{Draine2003:2003ARA&A..41..241D}. 
The size and the spatial distribution of dust thermal emission are essential to locate and characterize galaxy dust-obscured star formation, that in high-z DSFGs occurs in the form of intense bursts \citep[typical SFRs are of the order of $100-1000$ M$_\odot$ yr$^{-1}$, see e.g.][]{CaseyNarayananCooray2014:2014PhR...541...45C}.

We analyse the public ALMA archival maps containing our 11 DSFGs and we select the ones that have the best spatial resolution ($\Delta\theta\lesssim1$ arcsec) in the wavelength range $\lambda_{\rm obs}\sim 500\mu$m $-$ 3 mm (corresponding to the frequency range $\nu_{\rm obs}\sim 100-600$ GHz). Almost all of the best spatially-resolved continuum maps are in ALMA B6 and B7 that, indeed, constitute a very good compromise between spatial resolution and the sampled wavelength. This allow us to provide the most homogeneous information on dust continuum emission. Just for AzTEC.GS22 the image at the best spatial resolution is in B9.

In Tab. \ref{tab:continuum} we list the continuum flux densities (S$_\nu$) and sizes ($\theta_\nu$) of our sources. We estimate the flux density errors as:
\begin{equation}
    eS_\nu = \sqrt{({\rm rms})^2+(0.1\times S_\nu)^2}
\end{equation}
that is the quadratic sum of the ALMA continuum map rms and a conservative estimation of flux calibration accuracy, i.e. $\sim10$\% for the ALMA bands of our interest: B6, B7 and B9.

We measure the source sizes performing a 2D Gaussian fit of the source emission on the ALMA map by using the task \textit{imfit} embedded in the CASA viewer (CASA release 5.4.0-70). Some of the detections (6 out of 11) are not resolved in the corresponding ALMA map: in such a case we indicate the synthesized beam size (labelled with the apex \textit{sb}) that provides an upper limit on the source size. The level of noise did not allow us to extract the sizes from a fit in the visibility domains. 

We derive the linear circularized size of galaxy (sub-)millimeter emission by using the following expression:
\begin{equation}\label{eq_sizes}
 r_{\rm circ}(\nu)\,[{\rm kpc}] = \frac{a}{2} \,[{\rm arcsec}]\,\, \sqrt{\frac{b}{a}} \,\,c\,[{\rm kpc}/{\rm arcsec}]
\end{equation}
where $a$ and $b$ are the major and minor projected axes; $c$ is the angular-to-linear conversion factor, which depends on redshift and cosmology \citep[see Tab. 12 in][]{Pantoni2021:2021MNRAS.504..928P}; $a$ and $b$ are listed in Tab. \ref{tab:continuum} ($\theta_\nu=a_\nu\times b_\nu$). The resulting circularized ALMA sizes ($\rm{r_{circ}^{ALMA}}$, in kpc) are listed in Tab. \ref{tab:continuum}. When the source is not (entirely) resolved we indicate just an upper limit on its size (labelled with the symbol "$<$"), given by the circularized synthesized beam in kpc.

\begin{figure*}
    \centering
    \includegraphics[width=2.\columnwidth]{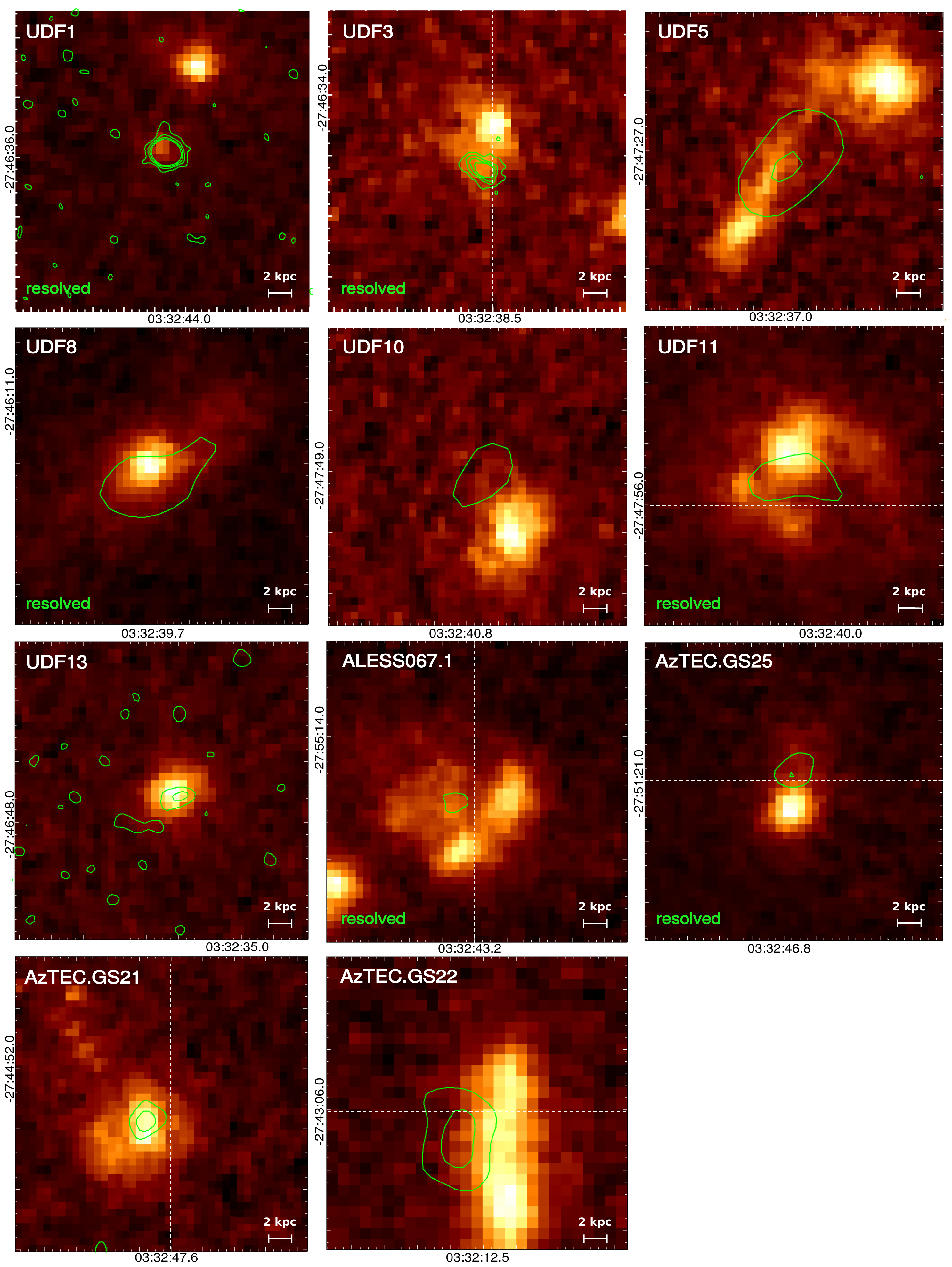}
    \caption{Postage stamps of 2.5 arcsec $\times$ 2.5 arcsec. ALMA continuum contours at $[1, 2, 3, 4]\times2.5\sigma$ (green solid lines) are overlaid on F160W HST/WFC3 images (corrected for astrometry). Resolved ALMA sources are labelled as such in the bottom left corner of the panel. In the x-axis and y-axis we show the Right Ascension (RA) in h:m:s and the Declination (DEC) in deg:arcmin:arcsec, respectively, for a reference point on the map.}
    \label{fig:cont}
\end{figure*}

In Fig. \ref{fig:cont} we show the ALMA continuum contours (at $[1,2,3,4]\times2.5\sigma$) overlapping the HST (H$_{160}$) image of the galaxies. Due to the low spatial resolution of \textit{Chandra} X-ray map \citep[][]{Luo2017:2017ApJS..228....2L}, we do not show galaxy X-ray emission on the HST maps. We just note that the peak of X-ray emission overlaps both the optical and (sub-)millimeter galaxy counterparts.
We corrected the HST images astrometry using the position of Gaia sources \citep[][]{Gaia2016:2016A&A...595A...2G} that are located in the same field of our DSFGs, finding a mean error of $\sim[+2,-10]\times10^{-5}$ degrees ($\Delta_{\rm RA}\sim70$ mas, $\Delta_{\rm DEC}\sim-360$ mas), in agreement with the astrometric error between VLA and HST images shown by \citet[][]{Rujopakarn2016:2016ApJ...833...12R}, i.e. 
$\Delta_{\rm RA}=+80\pm110$ mas, $\Delta_{\rm DEC}=-260\pm130$ mas \citep[see also e.g.][]{Dunlop2017:2017MNRAS.466..861D, Elbaz2018:2018A&A...616A.110E}. The astrometric correction on AzTEC.GS22 HST H$_{160}$ image is of $\sim[+15,+5]\times10^{-5}$ degrees, while we do not apply any correction on AzTEC.GS25 optical coordinates. After the correction, the bulks of the stellar and dust emission coincide within the uncertainties of the astrometric correction ($\sim100$ mas) and the beam resolution, for most of the galaxies. An eventual remaining shift tells us that the peaks of optical and (sub-)millimeter emission are not exactly coincident due to dust obscuration of stellar light.

\section{Spectroscopic emission}\label{sec:spectroscopicalemission}

CO transitions can be used to trace the molecular gas phase inside the galaxy and derive its mass, density and kinetic temperature \citep[e.g.][]{Yang2017:2017A&A...608A.144Y, Joblin2018:2018A&A...615A.129J,Boogaard2020:2020ApJ...902..109B}. A CO spectral line energy distribution (SLED) peaking at $J>3$ is typically considered an evidence of shocks and/or nuclear activity, originating from the so called X-ray Dominated Regions \citep[XDRs, see][]{Vallini2019:2019MNRAS.490.4502V}, while the low-J lines are more commonly associated to star-formation \citep[e.g.][]{Pozzi2017:2017MNRAS.470L..64P, Mingozzi2018:2018MNRAS.474.3640M,Carniani2019:2019MNRAS.489.3939C}, originating in the Photo-Dissociation Regions \citep[PDRs, see][]{Hollenbach1999:1999RvMP...71..173H}.  
Therefore, CO lines can be used to characterize both the ongoing star-forming burst \citep[such as the SFR, see e.g.][]{Bayet2009:2009MNRAS.399..264B} and the impact of the activity of the central SMBH on the host galaxy \citep[][]{Cicone2014:2014A&A...562A..21C}, giving some hints, in particular, on the possible connection between high-z Sub-millimeter Galaxies (SMGs) and AGN hosts \citep[e.g.][]{Sharon2016:2016ApJ...827...18S}, even if the CO SLED is not entirely sampled.  

\begin{figure}
    \centering
    \includegraphics[width=0.93\columnwidth]{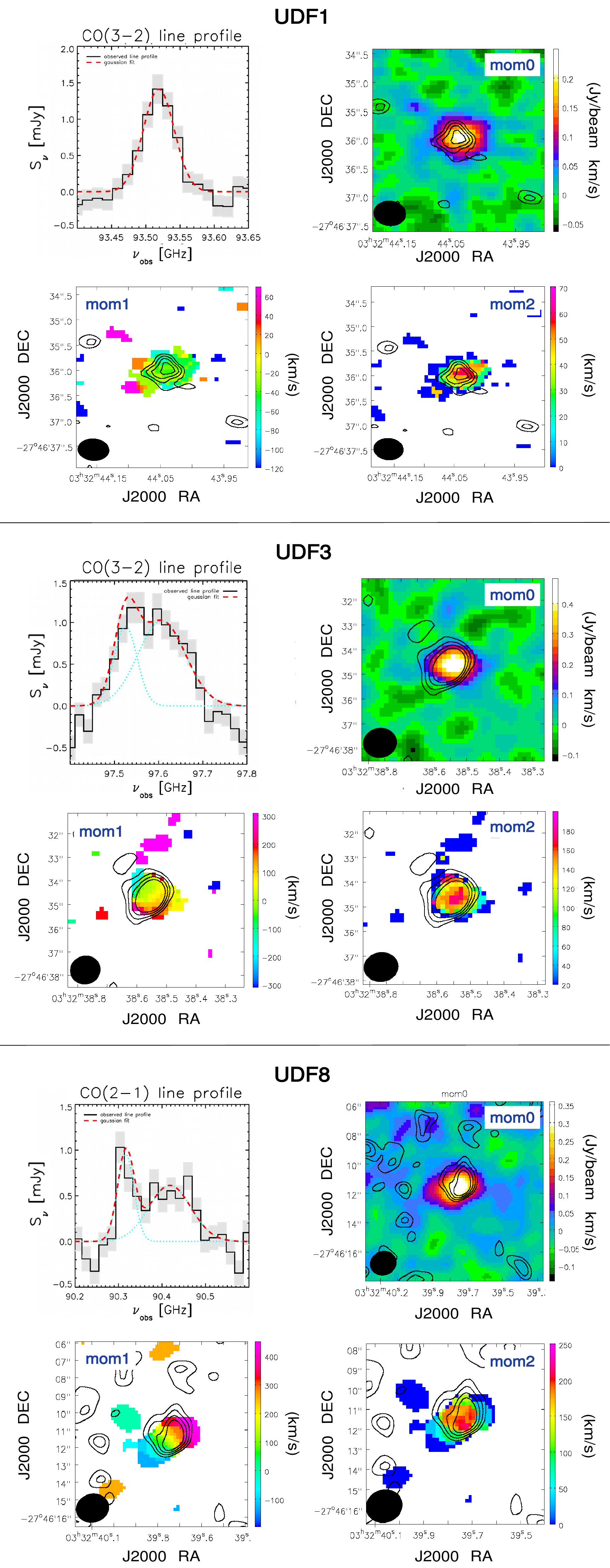}
    \caption{CO lines and spectral analysis for UDF1, UDF3 and UDF8 (from top to bottom). The panels show: the observed spectral line profile (black solid line), the corresponding uncertainties on flux density (gray shaded area) and the (total) best Gaussian fit (dashed red line; in case of double peaked profile the two components are in cyan); the maps of the spectral line distribution momenta (mom0, mom1 and mom2) overlapped by the continuum contours at $[2,3,4,5]\times\sigma$ detected at the same frequency (ALMA B3). Black filled ellipse in the bottom left corners indicates the beam size.}
    \label{fig:panel_CO_3sources}
\end{figure}

\begin{figure*}
    \centering
    \includegraphics[width=2.\columnwidth]{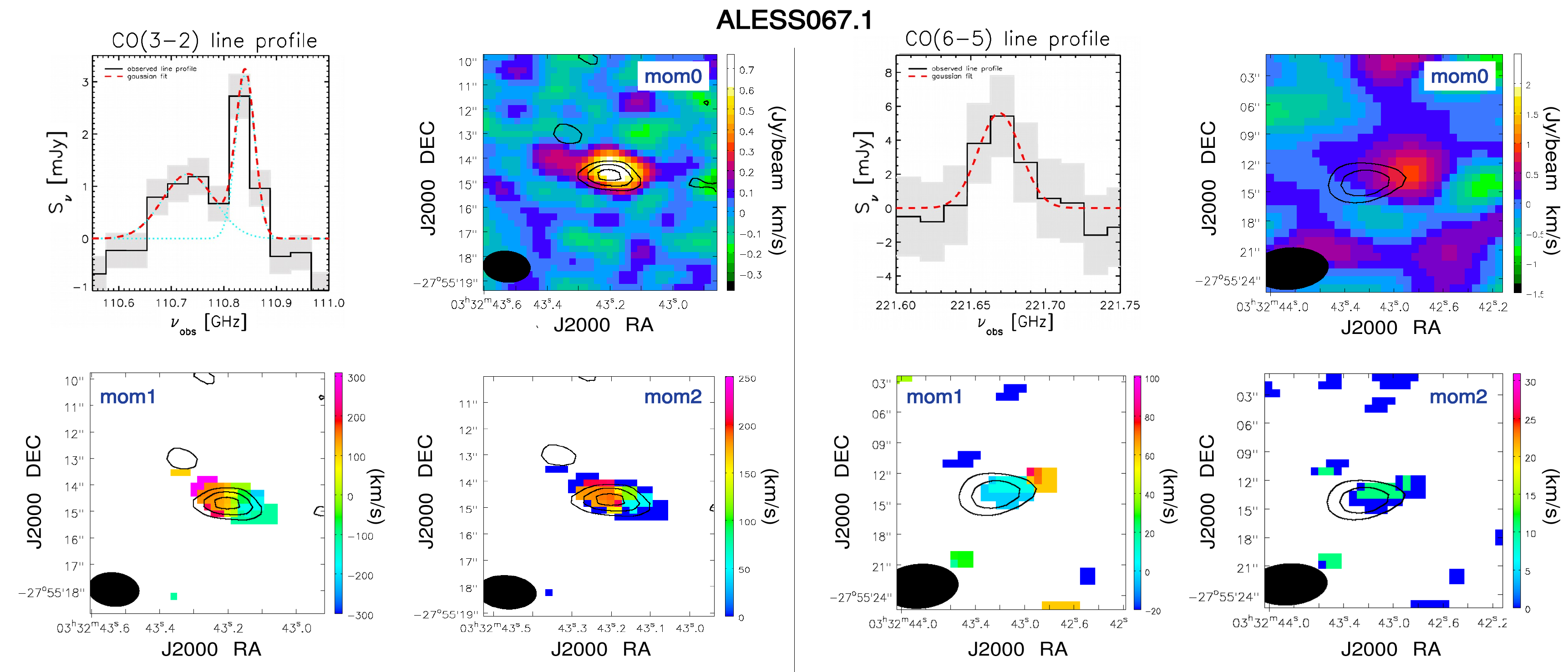}
    \caption{ALESS067.1 CO lines, i.e. $J=3$ and $J=6$, in the order. The panels show: the observed spectral line profile (black solid line), the corresponding uncertainties on flux density (gray shaded area) and the (total) best Gaussian fit (dashed red line; in case of double peaked profile the two components are in cyan); the maps of the spectral line distribution momenta (mom0, mom1 and mom2) overlapped by the continuum contours at $[2,3,4,5]\times\sigma$ detected at the same frequency (ALMA B3 and ALMA B6, respectively). Black filled ellipse in the bottom left corners indicates the beam size.}
    \label{fig:panel_CO_ALESS0671}
\end{figure*}

In the following we present the five CO $J>1$ line detections that we found for UDF1, UDF3, UDF8 and ALESS067.1 in the ALMA Archive. Images native spectral resolution in most cases was too small to allow the CO line profile to be clearly detected. In such cases we perform a rebin (using the CASA task \textit{imrebin}\footnote{In CASA, \textit{imrebin} performs an average over the binned quantities, in our cases the spectral channels.}) of the line channels. It consents to reduce the noise and boost the source signal by averaging among consecutive channels. We perform a $\times15$ rebin on UDF1 ALMA data cube and a $\times5$ rebin on ALESS067.1 ALMA (Project Code 2016.1.00564.S) data cube. We use instead data cubes of UDF3, UDF8 and ALESS067.1 ALMA (Project Code: 2019.2.00246.S) as available in the Archive with clean spectral resolution as indicated in Tab. \ref{tab:COemission}. In Figs. \ref{fig:panel_CO_3sources} and \ref{fig:panel_CO_ALESS0671} we show the observed spectral line profiles with their best Gaussian fits and the maps of spectral line distribution momenta (0, 1 and 2). CO intensity maps are overlapped by dust continuum contours. The angular and linear circularized sizes ($\theta_\nu$ and $r_{\rm CO}$) are listed in Tab. \ref{tab:COemission}. Since they all appear unresolved to a 2D Gaussian fit, we provide: an upper limit on UDF1 CO emission size, given by the synthesized beam; a measure of UDF3, UDF8 and ALESS067.1 CO emission equal to the angular distance between the peaks of the two spectral line components. We perform the conversion from angular to linear size exploiting Eq. \eqref{eq_sizes}. Then, we measure CO line intensity on mom0 maps and we list the corresponding values in Tab. \ref{tab:COemission}.

In Tab. \ref{tab:COemission} we list also the CO transition observed for each source and the central observed frequency $\nu_{\rm obs}$ that we use to compute the source redshift z (Tab. \ref{tab:galaxy_properties}). We note that most of the lines (CO($3-2$) for UDF3 and ALESS067.1; CO($2-1$) for UDF8; see Figs. \ref{fig:panel_CO_3sources} and \ref{fig:panel_CO_ALESS0671}) are characterized by an asymmetric double-peaked CO line profile, that could suggest that we are observing a tilted disc of rotating molecular gas or a molecular outflow produced by the central engine. Both the explanations are consistent with velocity maps (mom1), that are characterized by null central line (i.e., e.g. the rotation axis), and velocity dispersion maps (mom2), that are peaked along the same line. However, the width of the large components does not exceed 400 km/s, favouring the first scenario. Merging events seem to be less probable given the undisturbed appearance of the velocity  dispersion on the mom2 maps. 
UDF1 CO line profile is very narrow (FWHM $\sim$ 170 km/s) and show just a peak: it means that the galaxy cold gas component is actually not rotating or that we are looking at the galaxy \textit{face-on}. The latter interpretation is actually consistent with the velocity dispersion map, that is peaked in the centre. These evidences are confirmed also by the velocity range spanned by the CO component in the velocity maps: while for UDF3, UDF8 and ALESS067.1 the CO emission reaches velocities of the order of a few hundreds km/s, the CO velocity in UDF1 does not exceed 60 km/s, resembling the intrinsic chaotic motion of the cold gas phase. 

Finally, we focus on the case of ALESS067.1 that has two CO lines detected ($J=3$, $J=6$; Tab. \ref{tab:COemission}). The velocity map of ALESS067.1 CO(3-2) (mom1; Fig. \ref{fig:panel_CO_ALESS0671}) does not show a clear null line, while the velocity dispersion map (mom2; Fig. \ref{fig:panel_CO_ALESS0671}) peaks in the centre and shows a tail towards the left upper corner. More likely this could be an indication of an early AGN driven molecular outflow or an evidence of interactions \citep[e.g.][claim that ALESS067.1 is actually the central dominant galaxy of a multiple system]{Targett2013:2013MNRAS.432.2012T}. However, in the latter scenario we expect the momenta to be much more disturbed. In case we are actually observing a AGN driven outflow, we do not expect it to significantly affect the galaxy properties and its star formation activity at the moment, since the line FWHM is $<400$ km/s. CO(6-5) maps are both null in the centre, while mom1 map shows a peak to the right upper corner, peaking at 80 km/s. This could trace either the warm star formation of the galaxy or the central engine activity. The low spatial resolution of the image does not allow to spatially compare the size of the two spectral emission, neither to understand the origin of the $J=6$ line (a nuclear origin implies a very compact size of the CO emission, while a stellar one implies a more extended distribution). 
We need imaging at higher spatial resolution and sensitivity to definitively disentangle the diverse scenarios. Further details on ALESS067.1 CO spectral line emission along with a possible interpretation in the context of galaxy-BH co-evolution are described in Appendix \ref{app:ALESS067.1}.

\subsection{Molecular gas mass}\label{sec:gasmass}

In the following we exploit the CO line intensity to derive the molecular hydrogen mass content of UDF1, UDF3, UDF8 and ALESS067.1. 

We derive CO line luminosity using the following conversion by \citet[][]{Solomon1997:1997ApJ...478..144S}:
\begin{equation}
    L'_{\rm CO}\, {\rm[K\,km\,s^{-1}\, pc]} = 3.27\times10^{7} \,I_{\rm CO}\, \nu_{\rm obs}^{-2} \, d_{\rm L}^2 \,(1+z)^{-3}
\end{equation}
where $I_{\rm CO}$ is measured in Jy km s$^{-1}$; $\nu_{\rm obs}$ is the observed central frequency of the line (measured in GHz) and $z$ is the corresponding redshift. The luminosity distance $d_{\rm L}$ in Mpc depends on redshift and cosmology \citep[they are listed in][their Tab. 6, first column]{Pantoni2021:2021MNRAS.504..928P}. In Tab. \ref{tab:CO_gas_mol} we list line luminosities. We compute the associated uncertainties by using error propagation theory. 

We convert CO $J>1$ luminosity into the equivalent ground state luminosity $L'_{\rm CO(1-0)}$ by using the relation:
\begin{equation}
    L'_{\rm CO(1-0)}=\frac{L'_{\rm CO(J-[J-1])}}{r_{\rm J1}}
\end{equation}
and assuming the CO excitation ladder by \citet[][]{Daddi2015:2015A&A...577A..46D}, i.e. $r_{31}=0.42\pm0.07$ and $r_{21}=0.76\pm0.09$. 

We derive the molecular hydrogen mass by using the relation:
\begin{equation}
    M_{{\rm H}_2} [{\rm M}_\odot]=\alpha_{\rm CO} \, L'_{\rm CO(1-0)}
\end{equation}
where $\alpha_{\rm CO}$ is the CO conversion factor in units of M$_\odot$ [K km s$^{-1}$ pc$^2$]$^{-1}$. We assume an $\alpha_{\rm CO}=3.6$ M$_\odot$ [K km s$^{-1}$ pc$^2$]$^{-1}$ \citep[e.g.][]{Daddi2015:2015A&A...577A..46D, Decarli2016II:2016ApJ...833...70D}. The resulting molecular hydrogen masses are listed in Tab. \ref{tab:CO_gas_mol}. Uncertainties are computed with errors propagation theory.

We note that ALESS067.1 $H_2$ content (M$_{H_2}\sim1.7\times10^{11}$ M$_\odot$; see Tab. \ref{tab:CO_gas_mol}) is consistent with the molecular hydrogen mass derived from the CO(1-0) line luminosity measured with the Australian Telescope Compact Array (ATCA) by \citet[][]{Huynh2017:2017MNRAS.467.1222H}, i.e. $L'_{\rm CO(1-0)}=(9.9\pm1.8)\time10^{10}$ K km s$^{-1}$ pc$^2$, and assuming an $\alpha_{\rm CO}=1.8$ M$_\odot$ [K km s$^{-1}$ pc$^2$]$^{-1}$ \citep[i.e. M$_{H_2}\sim1.8\times10^{11}$ M$_\odot$; see][]{Chen2017:2017ApJ...846..108C}. The latter conversion factor is often thought to be preferable for compact SMGs \citep[e.g.][]{Chen2017:2017ApJ...846..108C, Elbaz2018:2018A&A...616A.110E, Carilli2013:2013ARA&A..51..105C}. However, due to the large uncertainties on $\alpha_{\rm CO}$, we do not favour one or the other value and we suggest just to re-scale the resulting molecular gas in Tab. \ref{tab:CO_gas_mol} by a factor $1.8/3.6=0.5$. The presence of the direct measurement of the CO(1-0) line luminosity for ALESS067.1 by \citet[][]{Huynh2017:2017MNRAS.467.1222H} allow us to compute $r_{31}$ and $r_{61}$. We calculate a line luminosity ratio of $r_{31}=L'_{\rm CO(3-2)}/L'_{\rm CO(1-0)}=0.20\pm0.07$ and $r_{61}=L'_{\rm CO(6-5)}/L'_{\rm CO(1-0)}=0.12\pm0.07$. We note that the $r_{31}$ we obtain is smaller than the one measured by \citet[][]{Daddi2015:2015A&A...577A..46D}, indicating a more excited CO SLED for ALESS067.1 if compared to the normal (near-IR selected BzK) star-forming disk galaxies at $z=1.5$ studied by \citet[][]{Daddi2015:2015A&A...577A..46D}. 
Typical values of $r_{61}$ are $\gtrsim0.2$ for SMGs and even higher for QSOs \citep[e.g., ][extrapolations from the latter two]{Bothwell2013:2013MNRAS.429.3047B, Carilli2013:2013ARA&A..51..105C, Daddi2015:2015A&A...577A..46D}. The sightly lower value we find is probably due to sensitivity limit that makes us miss the outskirts of the CO emission, coupled with low resolution that causes the flux to be distributed on a larger image area, while for high-J it is probably mostly concentrated in the central region.

\begin{table}
 \centering 
 \caption{CO analysis: $L_{\rm CO}$ and M$_{H_2}$ for UDF1, UDF3 UDF8 and ALESS067.1, by assuming an $\alpha_{\rm CO}=3.6$ K km pc$^2$ s$^{-1}$ M$_{\rm \odot}^{-1}$.}\label{tab:CO_gas_mol}
  \begin{tabular}{lcccc}
  \hline
 $\mathbf{ID}$   & $\mathbf{z_{CO}}$  & $\mathbf{J^{up}}$  & $\mathbf{L_{CO}}$  & $\mathbf{M_{H_2}}$ \\
 &  &  & [${10}^8$ K km s$^{-1}$ pc$^2$] &[$10^{10}$ M$_\odot$] \\
 \hline
 UDF1 & 2.698 & 3 &  $31\pm3$ & $2.6\pm0.7$ \\
 UDF3 & 2.543 & 3 &  $170\pm10$ & $15\pm3$  \\
 UDF8 & 1.5490 & 2 & $122\pm9$ & $5.8\pm1.1$ \\
 ALESS067.1 & 2.1212 & 3 & $196\pm31$ & $16.8\pm5.4$ \\
 & 2.1212 & 6 & $123\pm45$& $-$\\
  \hline
  \end{tabular}
 $H_2$ mass of ALESS067.1 is calculated by exploiting the transition with lower J$^{\rm up}$.
\end{table}

\section{Discussion}\label{sec:discussion}

In this Section we combine our broad-band spatially-resolved and spectral analyses with the results obtained from galaxy SED fitting by \citet[][]{Pantoni2021:2021MNRAS.504..928P} and additional information collected from literature and multi-wavelength images from public catalogs, in order to include the whole sample in a broad context of galaxy formation and evolution.

\begin{table*}
 \centering
 \caption{Integral properties of the 11 DSFGs in our sample by \citet[][]{Pantoni2021:2021MNRAS.504..928P}. We list (in the order): galaxy ID; spectroscopic redshift (references in Tab. \ref{tab:galaxy_properties}); stellar mass (M$_\odot$); Star Formation Rate (SFR); burst age ($\tau_\star$); IR luminosity (L$_{\rm IR}$); AGN fraction in the IR ($f_{\rm AGN}$); dust mass (M$_{\rm dust}$); gas mass (M$_{\rm gas}$); gas depletion timescale ($\tau_{\rm depl}$); $2-10$ keV luminosity (L$_{\rm X}$); X-ray dominant component (class X); presence of an AGN (AGN) by \citet[][]{Luo2017:2017ApJS..228....2L}.}
 \label{tab:summary_integral}
  \resizebox{2.1\columnwidth}{!}{
\hspace{-0.5cm}
  \begin{tabular}{llccccccccccc}
  \hline
  \textbf{ID}  & $\mathbf{z_{spec}}$ &  $\mathbf{M_\star}$ & \textbf{SFR} & $\mathbf{\tau_\star}$ & $\mathbf{L_{IR}}$ & $\mathbf{f_{AGN}}^{(1)}$ & $\mathbf{M_{dust}}^{(2)}$ & $\mathbf{M_{gas}}^{(3)}$ & $\mathbf{\tau_{depl}}$ & $\mathbf{L_{X}}$ & \textbf{class X} & \textbf{AGN} \\
   &  & [$10^{10}$ M$_\odot$] & [M$_\odot$ yr$^{-1}$] & [Myr] & [$10^{12}$ L$_\odot$] & [\%] & [$10^{8}$ M$_\odot$] & [$10^{10}$ M$_\odot$] & [Myr] &[$10^{42}$ erg s$^{-1}$] &  & \\
  \hline
   UDF1 & 2.698 & $8\pm1$ & $352\pm18$ & $334\pm58$ & $3.5\pm0.2$ & 6 & $5.6\pm0.2$ & $3^{+3}_{-1}$ & 85 & 40.2 & AGN & \cmark\\
   UDF3 & 2.543 & $9\pm1$ & $519\pm38$ & $234\pm47$ &  $4.9\pm0.3$ & 0.2 & $4.0\pm1.6$ & $4^{+4}_{-2}$ & 77 & 1.8 & galaxy & \cmark\\
   UDF5 & 1.759 & $2.4\pm0.3$ & $85\pm6$ & $404\pm85$ & $0.77\pm0.04$ & $-$ & $4.6\pm2.2$ & $1.3^{+2.0}_{-0.4}$ & 153 & $-$ & $-$ & \textbf{nd}\\
   UDF8 & 1.549 & $6.5\pm0.3$ & $100\pm5$ & $992\pm50$ & $1.10\pm0.06$ & 14 & $2.4\pm1.3$ & $1.0^{+1.0}_{-0.5}$ & 100 & 36.3 & AGN & \cmark\\
   UDF10 & 2.086 & $2.5\pm0.3$ & $41\pm5$ & $917\pm137$ & $0.41\pm0.05$ & 1 & $1.9\pm1.3$ & $0.6^{+0.4}_{-0.3}$ & 146 & 0.6 & galaxy & \xmark \\
   UDF11 & 1.9962 & $6.4\pm0.9$ & $241\pm19$ & $380\pm82$ & $2.2\pm0.2$ & 0.5 & $1.46\pm0.66$ & $0.6^{+0.7}_{-0.3}$ & 25 & 1.7 & galaxy & \xmark \\
   UDF13 & 2.497 & $6.5\pm1.4$ & $111\pm17$ & $879\pm149$ & $1.2\pm0.2$ & 0.8 & $1.20\pm0.68$ & $0.5^{+0.5}_{-0.2}$ & 45 & 2.1 & galaxy & \cmark\\
   ALESS067.1 & 2.1212 & $29\pm3$ & $487\pm24$ & $903\pm100$ &$5.4\pm0.3$ & 0.4 & $10\pm4$ & $8^{+8}_{-4}$ & 164 & 3.8 & galaxy & \cmark\\
   AzTEC.GS25 & 2.292 & $8\pm2$ & $401\pm20$ & $290\pm88$ & $3.9\pm0.2$ & 1 & $13.6\pm8.2$ & $4^{+4}_{-2}$ & 100 & 6.1 & galaxy & \cmark \\ 
   AzTEC.GS21 & 1.910 & $18\pm2$ & $360\pm18$ & $746\pm105$ & $3.9\pm0.2$ & 0.3 & $5.8\pm1.4$ & $5^{+5}_{-2}$ & 139 & 1.7 & galaxy & \cmark \\
   AzTEC.GS22 & 1.794 & $5.7\pm0.5$ & $91\pm5$ & $940\pm74$ & $1.01\pm0.06$ & $-$ & $14\pm8$ & $5^{+5}_{-2}$ & 550 & $-$ & $-$ & \textbf{nd} \\
  \hline
  UDF2 & $2.6961^{(a)}$ & $13\pm3^{(a)}$ & ${187^{+35}_{-16}}^{(a)}$ & $-$ & $-$ & $-$ & ${7.8^{+1.2}_{1.0}}^{(c)}$ & Tab. \ref{tab:summary_resolved} & Tab. \ref{tab:summary_resolved} & $-$ & $-$ & \textbf{nd} \\
  \hline
  \end{tabular}
  }
 The AGN fractions (1) reported above are the ones inferred by using the MIR-X-ray correlation by \citet[][]{Asmus2015:2015MNRAS.454..766A}; cf. \citet[][their Sect. 4.4.2]{Pantoni2021:2021MNRAS.504..928P}. The dust masses (2) listed above include the correction by \citet[][]{Magdis2012;2012ApJ...760....6M} of a factor $\sim2$. Gas masses (3) are derived from the dust continuum flux at $\lambda=850\,\mu$m, following \citet[][cf. their Fig. 1]{Scoville2016:2016ApJ...820...83S}; see also \citet[][their Sect.4.3 and in particular their Eq. 5]{Pantoni2021:2021MNRAS.504..928P}. In column \textbf{AGN} we take into consideration any evidence of the presence of an AGN, as reported by \citet[][]{Luo2017:2017ApJS..228....2L}: \cmark $\,$ means that the galaxy has a signature of AGN; \xmark $\,$ means that the object is classified as a normal star-forming galaxy; \textbf{nd} stands for \textit{not detected}. For reference we list the same available information for UDF2 by \citet[][]{Boogaard2019:2019ApJ...882..140B}$^{(a)}$, \citet[][]{Rujopakarn2019:2019ApJ...882..107R}$^{(b)}$ and \citet[][]{Kaasinen2020:2020ApJ...899...37K}$^{(c)}$.
\end{table*}

\begin{table*}
 \centering
 \caption{Spatially-resolved properties of the 11 DSFGs in our sample and information derived from CO emission lines. We list (in the order): galaxy ID; spectroscopic redshift (references in Tab. \ref{tab:galaxy_properties}); optical (H$_{160}$) circularized radius ($r_{\rm H}$); thermal dust circularized radius by ALMA continuum ($r_{\rm ALMA}$); CO circularized radius ($r_{\rm CO}$); optical-to-ALMA size ratio (r$_{\rm H}/$r$_{\rm ALMA}$); optical-to-CO size ratio (r$_{\rm H}/$r$_{\rm CO}$); HST morphology in the filter H$_{160}$; CO transition upper level ($J^{\rm up}$); CO line luminosity (L$_{\rm CO}$); $H_2$ mass (M$_{H_2}$); gas depletion timescale ($\tau_{\rm depl}$).}
 \label{tab:summary_resolved}
  \resizebox{2.1\columnwidth}{!}{
\hspace{-0.5cm}
  \begin{tabular}{llccccccccccc}
  \hline
  \textbf{ID}  & $\mathbf{z_{spec}}$   &  $\mathbf{r_{H}}$ & $\mathbf{r_{ALMA}}$ & $\mathbf{r_{CO}}$ & $\mathbf{r_{H}/r_{ALMA}}$ & $\mathbf{r_{H}/r_{CO}}$ & $\mathbf{r_{CO}/r_{ALMA}}$ & \textbf{morphology} & $\mathbf{J^{up}}$ & $\mathbf{L_{CO}}$ & $\mathbf{M_{H_2}}$ & $\mathbf{\tau_{depl}}$\\
   &  &  [kpc] & [kpc] & [kpc] & & & & $H_{160}$ & & [${10}^8$ K km s$^{-1}$ pc$^2$] & [$10^{10}$ M$_\odot$] & [Myr]\\
  \hline
   UDF1 & 2.698 & 2.6 & 0.46 & $<3.3$ & 5.7 & $-$ & $-$ & isolated & 3 & $31\pm3$ & $2.6\pm0.7$ & 74\\
   UDF3 & 2.543 & 1.6 & 0.62 & 1.35 & 2.6 & 1.2 & 2.2 & clumpy? & 3 & $170\pm10$ & $15\pm3$ & 289\\
   UDF5 & 1.759 & 2.3 & $<2.5$ & $-$ & $-$  & $-$ & $-$& clumpy & $-$ & $-$ & $-$ & $-$ \\
   UDF8 & 1.549 & 5.7 & 4.1 & 3.45 & 1.4 & 1.7 &0.9 & isolated & 2 & $122\pm9$ & $5.8\pm1.1$ & 580\\
   UDF10 & 2.086 & 2.0 & $<2.5$ & $-$ & $-$  & $-$ & $-$ & clumpy & $-$ &$-$ &$-$ &$-$ \\
   UDF11 & 1.9962 & 4.5 & 3.4 & $-$ & 1.3 & $-$  & $-$ & clumpy & $-$ & $-$ & $-$ & $-$\\
   UDF13 & 2.497 & 1.2 & $<0.65$ & $-$ & $-$  & $-$ & $-$ & isolated & $-$ & $-$ & $-$ & $-$\\
   ALESS067.1 & 2.1212 & 6.5 & 1.1 & 3.6 & 6.0 & 1.8 &3.3 & clumpy & 3 & $196\pm31$ & $16.8\pm5.4$ & 345\\
   &  & &  &  &  &  & & & 6 & $123\pm45$ & & \\
   AzTEC.GS25 & 2.292 & 1.8 & 1.2 & $-$  & 1.5 & $-$  & $-$ & isolated & $-$ & $-$ & $-$ & $-$\\ 
   AzTEC.GS21 & 1.910 & 3.7 & $<0.7$ & $-$ & $-$ & $-$  & $-$ & clumpy & $-$ & $-$ & $-$ & $-$\\
   AzTEC.GS22 & 1.794 & 3.2 & $<1.7$ & $-$ & $-$ & $-$ & $-$ & clumpy & $-$ & $-$ & $-$ & $-$\\
  \hline
  UDF2 & $2.6961^{(a)}$ & 2.5$^{(b)}$ & $0.6^{(*)}$ & $ 2.6^{(c)}$ & 4.3 & 0.97 & 4.3 & clumpy$^{(b)}$ & $3^{(a)}$ & $279\pm33^{(a)}$ & $23.9\pm4.9^{(a)}$ & $1300^{(a)}$ \\
  \hline
  \end{tabular}
  }
 For reference we list the same information for UDF2 by \citet[][]{Boogaard2019:2019ApJ...882..140B}$^{(a)}$, \citet[][]{Rujopakarn2019:2019ApJ...882..107R}$^{(b)}$ and \citet[][]{Kaasinen2020:2020ApJ...899...37K}$^{(c)}$.
 (*)\citet[][]{Rujopakarn2019:2019ApJ...882..107R} find a core component of 0.3 kpc and a disk component of 1.2 kpc (circularized radii). 
\end{table*}

In Tabs. \ref{tab:summary_integral} and \ref{tab:summary_resolved} we list the global astrophysical properties of the individual 11 DSFGs in our sample, while in Tab. \ref{tab:sample_median_value} we show their median values, with the corresponding first and third quartiles. For reference we compare the outcomes with a well studied $z\sim2$ DSFG (i.e. UDF2) that is not included in our sample, but presents a similar multi-band sampling of its SED and the same multi-wavelength spatially-resolved and spectral information \citep[see e.g.,][]{Boogaard2019:2019ApJ...882..140B, Rujopakarn2019:2019ApJ...882..107R, Kaasinen2020:2020ApJ...899...37K}.

\begin{figure}
    \centering
    \includegraphics[width=1.\columnwidth]{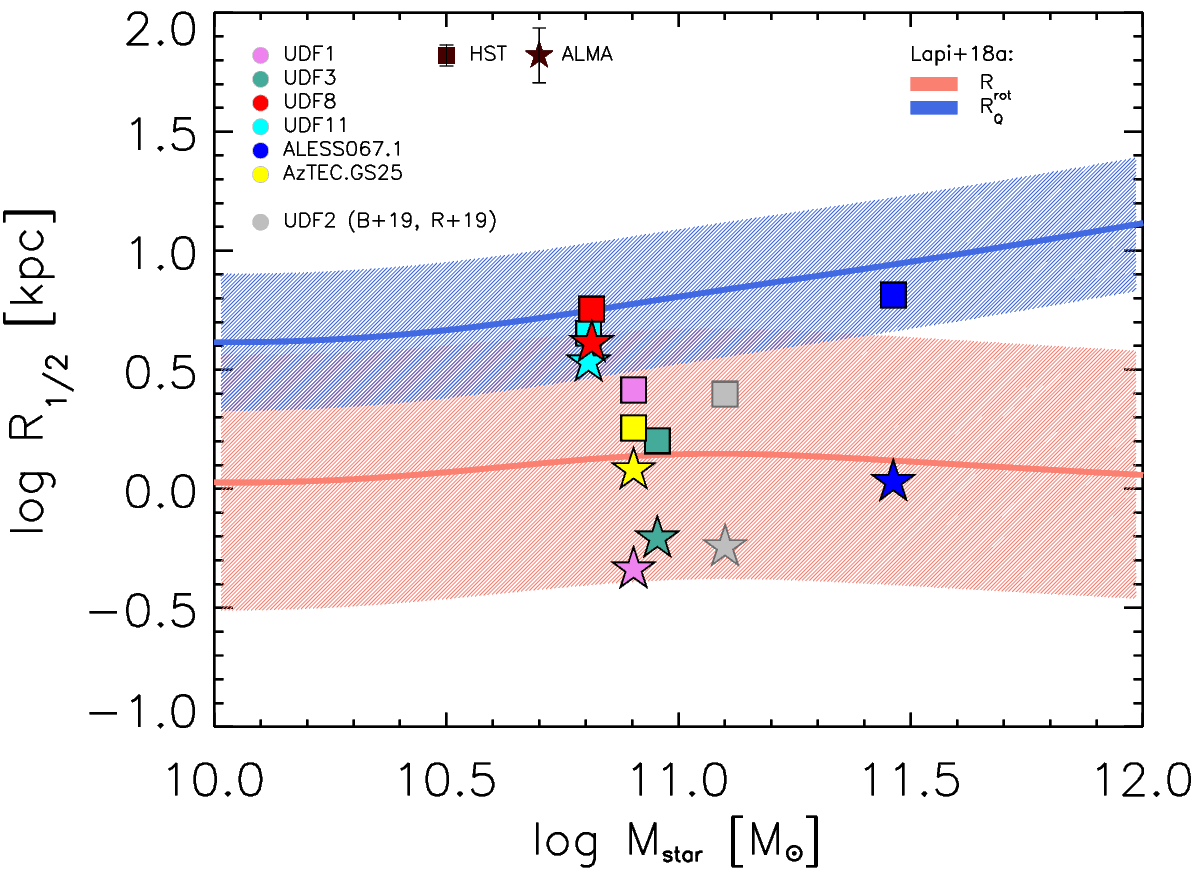}
    \caption{H$_{160}$ HST sizes (filled squares) and ALMA continuum sizes (filled stars) for the six galaxy that are resolved in the continuum ALMA maps: UDF1 (lilac); UDF3 (green); UDF8 (red); UDF11 (cyan); ALESS067.1 (cyan); AzTEC.GS25 (yellow). Gray filled symbols stand for UDF2 by \citet[][B+19]{Boogaard2019:2019ApJ...882..140B} and \citet[][R+19]{Rujopakarn2019:2019ApJ...882..107R}, that we show as an example of the typical outcomes from similar spatially-resolved and panchromatic studies on $z\sim2$ dusty star-forming galaxies. Blue and red shaded areas represent the predicted behaviour (with its $1\sigma$ scatter) of R$_{\rm Q}$ and R$_{\rm rot}$ by the theoretical scenario presented in \citet[][]{Lapi2018a:2018ApJ...857...22L}. Error bars, consistent with the scatter of the relations, have been omitted for clarity. Typical errors on HST and ALMA sizes are shown in the legend.}
    \label{fig:sizes}
\end{figure}

In Tab. \ref{tab:summary_resolved} we compute the ratio between the optical (H$_{160}$) circularized radius and the ALMA size, i.e. r$_{\rm H}/$r$_{\rm ALMA}$. We note that the optical size is typically equal or more extended than the (sub-)millimeter one.
This trend is actually observed in many recent works on high-z SMGs \citep[e.g.,][]{Barro2013:2013ApJ...765..104B, Barro2016I:2016ApJ...820..120B, Barro2016II:2016ApJ...827L..32B, Ikarashi2015:2015ApJ...810..133I, Simpson2015:2015ApJ...807..128S, Talia2018:2018MNRAS.476.3956T, Tadaki2020:2020ApJ...901...74T, Massardi2018:2018A&A...610A..53M}, and predicted by some theoretical models describing massive galaxy evolution, focusing in particular on the star forming progenitors of $z<1$ Early Type Galaxies (ETGs). 
In Fig. \ref{fig:sizes} we show the size-mass relation as it is predicted in the theoretical scenario by \citet[][]{Lapi2018a:2018ApJ...857...22L}, in terms of two typical radii: the stability radius to gas fragmentation R$_{\rm Q}$ ($\sim$ a few kpc for ETG star-forming progenitors at $z>1$), that is derived from the Toomre stability criterion \citep[see][their Eqs. 9 and 10]{Toomre1964:1964ApJ...139.1217T,  Lapi2018a:2018ApJ...857...22L}; the rotational radius R$_{\rm rot}$ ($\sim1$ kpc for ETG star-forming progenitors at $z>1$), for which the rotational support balances the gravitational pull of the inflowing gas \citep[][their Eqs. 16 and 17]{Lapi2018a:2018ApJ...857...22L}. 
The scenario predicts that the star-forming progenitors of local ETGs experience, at high-z, a dusty burst of star formation in the very central region of the galaxy, at radii $\lesssim$ R$_{\rm rot}$ (red solid line with its 1$\sigma$ scatter in Fig. \ref{fig:sizes}), that is traced by the size of dust continuum in the (sub-)millimeter band. 
At greater radii, i.e. R$_{\rm rot}$ $\lesssim r \lesssim$ R$_{\rm Q}$ (blue solid line with its 1$\sigma$ scatter in Fig. \ref{fig:sizes}) the star formation is less obscured by dust, so that UV/optical radiation from newly formed stars can partially emerge. This trend is present in all the six sources of our sample that are resolved both in the HST H$_{160}$ (optical sizes; filled squares in Fig. \ref{fig:sizes}) and ALMA continuum maps (millimeter sizes; filled stars in Fig. \ref{fig:sizes}), i.e., UDF1, UDF3, UDF8, UDF11, ALESS067.1 and AzTEC.GS25, for which the optical size is more extended than the millimeter one. The median optical-to-millimeter radii ratio, i.e. r$_{\rm H}/$r$_{\rm ALMA}=2.05$, may be just a lower limit of the real situation, since a significant part of the stellar emission could be lost, being completely absorbed by dust. Furthermore, the ratio r$_{\star}/$r$_{\rm dust}$ for our high-z objects behaves differently than for local samples of similar stellar mass \citep[e.g.,][]{Lang2019:2019ApJ...879...54L}, being typically larger than 1 for the former and smaller than 1 for the latter .

For reference, in Fig. \ref{fig:sizes} we show the position of UDF2 (gray filled symbols), that was detected in the ALMA 1.3 mm survey by \citet[][]{Dunlop2017:2017MNRAS.466..861D} at $z_{\rm spec}=2.6961$ \citep[][source ID: ASPECS-LP.3mm.07]{Boogaard2019:2019ApJ...882..140B}. We stress that the source is not included in our sample since it has no counterpart in the GOODS-MUSIC catalog by \citet[][]{Grazian2006:2006A&A...449..951G}, within the searching radius of 1 arcsec \citep[more details in][cf. their Sect. 2]{Pantoni2021:2021MNRAS.504..928P}. Its physical properties are very similar to our 11 DSFGs. It has a stellar mass M$_\star= (1.3\pm0.3)\times10^{11}$ M$_\odot$ and it forms stars at a rate of SFR $=187^{+35}_{-16}$ M$_\odot$ yr$^{-1}$ \citep[see Tab. \ref{tab:summary_integral} and][source ID: ASPECS-LP.3mm.07]{Boogaard2019:2019ApJ...882..140B}. 
Its spatially-resolved continuum emission, both in the optical (H$_{160}$) and in the (sub-)millimeter (ALMA B7), is analysed in the details by \citet[][]{Rujopakarn2019:2019ApJ...882..107R}. For the sake of consistency and to compare with our results, we measure the dust size on the same ALMA map (publicly available in the ALMA Archive; ALMA Project code: 2017.1.00001.S; Member ous: A001/X1288/X4c3) and calculate a circularized ALMA radius of $r_{\rm B7}\approx0.6$ kpc, while \citet[][]{Rujopakarn2019:2019ApJ...882..107R} measure a core component of $\approx 0.3$ kpc and a disc component of $\approx1.2$ kpc (circularized radii; see Tab. \ref{tab:summary_resolved}). The circularized H$_{160}$ radius, i.e. $r_{\rm H}\approx2.5$ kpc (Tab. \ref{tab:summary_resolved}), is obtained from the effective radius by \citet[][]{Rujopakarn2016:2016ApJ...833...12R}, i.e. $r_e=3.71\pm0.06$ kpc by assuming the axes ratio by \citet[][]{vanderWel2012:2012ApJS..203...24V}.
As for the other sources in our sample, also the HST and ALMA sizes of UDF2 are in good agreement with the size-mass relation predicted by \citet[][]{Lapi2018a:2018ApJ...857...22L} and shown in Fig. \ref{fig:sizes}.  

On the one hand, this evidence further confirms that our approach may be applied to other high-z DSFGs with similar multi-band coverage of their SED and spectral information, in the case spatially-resolved images are available in the FIR and UV/optical rest-frame. On the other hand, it stresses again the great importance of multi-wavelength imaging at great spatial resolution in order to probe galaxy evolution.

\begin{table}
 \centering
 \caption{Median, first  and third quartiles of the following quantities (in the order): redshift ($z$), age of the burst ($\tau_\star$); SFR; stellar mass (M$_\star$); IR luminosity (L$_{\rm IR}$); dust mass (M$_{\rm dust}$); molecular gas mass (M$_{\rm gas}$); depletion time ($\tau_{\rm depl}$); AGN fraction in the IR domain ($f_{\rm AGN}$); $2-10$ keV luminosity (L$_{\rm X}$); HST, ALMA and CO sizes (r$_{\rm H}$, r$_{\rm ALMA}$ and r$_{\rm CO}$); optical-to-ALMA size ratio (r$_{\rm H}/$r$_{\rm ALMA}$); optical-to-CO size ratio (r$_{\rm H}/$r$_{\rm CO}$); CO-to-ALMA size ratio (r$_{\rm CO}/$r$_{\rm ALMA}$). For reference, in the last column we list the values measured for UDF2 by\citet[][]{Boogaard2019:2019ApJ...882..140B}$^{(a)}$, \citet[][]{Rujopakarn2019:2019ApJ...882..107R}$^{(b)}$ and \citet[][]{Kaasinen2020:2020ApJ...899...37K}$^{(c)}$. }\label{tab:sample_median_value}
 \resizebox{1.\columnwidth}{!}{
\hspace{-0.5cm}
  \begin{tabular}{llcccc}
  \hline
   & & \textbf{Median} & $\mathbf{1^{st}}$ \textbf{quartile} & $\mathbf{3^{rd}}$ \textbf{quartile} & \textbf{UDF2}\\
 \hline
   z & &2.086 & 1.794 & 2.497 & 2.6961\\
   SFR & [M$_\odot$ yr$^{-1}$] &241 & 91 & 401 & 187\\
   $\tau_\star$ & [Myr] & 746 & 334 & 917 & $-$\\
   M$_{\star}$ & [$10^{10}$ M$_\odot$] & 6.5 & 5.7 & 9 & 13\\
   L$_{\rm IR}$ & [$10^{12}$ L$_\odot$] & 2.2 & 1.01 & 3.9 & $-$\\
   M$_{\rm dust}$ & [$10^{8}$ M$_\odot$] & 4.6 & 1.9 & 10 & 7.8\\
   M$_{\rm gas}$ & [$10^{10}$ M$_\odot$] & 4.0 & 0.6 & 5.8 & 23.9\\
   $\tau_{\rm depl}$  & [Myr]& 146 & 74 & 345 & 1300\\ 
   f$_{\rm AGN}$ & [\%] &0.8 & 0.4 & 1 & $-$\\
   L$_{\rm X}$ & [$10^{42}$ erg s$^{-1}$] & 1.7 & 2.1 & 6.1 & nd\\
   r$_{\rm H}$ & [kpc] & 2.6 & 1.8 & 4.5 & 2.5\\
   r$_{\rm ALMA}$ & [kpc] & 1.15 & 0.6 & 3.4 & 0.6\\
   r$_{\rm CO}$ & [kpc] & 3.45 & 1.35 & 3.6 & 2.6\\
   r$_{\rm H}/$r$_{\rm ALMA}$ &  & 2.05 & 1.4 & 5.7 & 4.3\\
   r$_{\rm H}/$r$_{\rm CO}$ & & 1.7 & 1.2 & 1.8 & 0.97\\
   r$_{\rm CO}/$r$_{\rm ALMA}$ & & 2.2 & 0.9 & 3.3 & 4.3\\
  \hline
  \end{tabular}
  }
  To compute the gas mass value reported above we preferred the measurements from CO spectral lines (Tab. \ref{tab:summary_resolved}) rather than the ones from dust continuum (Tab. \ref{tab:summary_integral}).
\end{table}

More than half of the objects in our sample (i.e., UDF5, UDF10, UDF11, ALESS067.1, AzTEC.GS21, AzTEC.GS22), as well as UDF2 (see Tab. \ref{tab:summary_resolved}), shows a clumpy morphology in the H$_{160}$ image. This may be suggestive of interactions within a radius $\lesssim$ a few tens of kpc from the ALMA counterpart. However, the optical clumps may just indicate that some emitting areas in the star-forming regions are affected more than others by dust extinction (indeed they are often indicated as \textit{star-forming clumps}). In order to discern the most likely scenario, it is essential to take into consideration galaxy multi-wavelength emission and its spatial distribution at comparable resolutions. In Appendix \ref{appendix} we provide an insight on this topic, for each galaxy in our sample.

Furthermore atomic and molecular spectral lines provide essential information to investigate the mechanisms triggering star formation in galaxies and galaxy-BH co-evolution. CO lines are very well recognized as tracers of the cold (i.e., molecular) gas phase and the analysis of their resolved emission, both in space and frequency, allow to study gas kinematics and physical conditions and to measure its content in mass. 
We identified some differences in the molecular gas masses estimated by using the dust optically-thin continuum and CO lines, that obviously imply diverse depletion timescales $\tau_\star$ (cf. Tab. \ref{tab:summary_integral} and Tab. \ref{tab:summary_resolved}). These differences can be traced back to diverse dust and CO sizes, the latter being typically larger than the former and its emission almost optically thick (especially in case of low-J CO lines). As such, gas masses inferred from these measurements may sample different regions or components in the galaxy \citep[e.g.,][]{Riechers2011:2011ApJ...739L..31R, Hodge2015:2015ApJ...798L..18H, Spilker2015:2015ApJ...811..124S, Decarli2016II:2016ApJ...833...70D}.

For the 4 sources in our sample with detected CO lines (i.e., UDF1, UDF3, UDF8, ALESS067.1), and also for UDF2, the cold gas emission is equally or more extended then the ALMA B7 and B6 continuum (typically r$_{\rm CO}/$r$_{\rm ALMA}\gtrsim1$), while it is equal or more compact than the optical emission (see Tab. \ref{tab:summary_resolved}). This is consistent with the size evolution scenario presented in \citet[][]{Lapi2018a:2018ApJ...857...22L} according to which
the CO emission traces the rotating cold gas phase that, while inflowing towards the central regions of the galaxy, fuels both the mildly-obscured star formation at larger radii (traced by the emission in the optical) and the highly-obscured star formation in the innermost regions (sampled by the dust continuum in the (sub-)millimeter).

Furthermore, multi-wavelength observations at high spectral and spatial resolution can reveal the presence of nuclear activity. In particular, structure in the CO emission may trace molecular outflows associated with AGN feedback. For example, the X-ray emission properties combined with the double peak CO line profile of UDF8, are strongly suggestive of the presence of nuclear driven ouflows. This interpretation is also consistent with the age of UDF8 and its SFR (i.e., $\tau_\star\sim1$ Gyr and SFR $\sim 100$ M$_\odot$ yr$^{-1}$; cf. Tab. \ref{tab:summary_integral}), in the galaxy-BH co-evolution scenario by \cite{Mancuso2016b:2016ApJ...833..152M, Mancuso2017:2017ApJ...842...95M}.
A more detailed analysis for UDF8 is given in the Appendix \ref{appendix}.

\section{Summary and conclusions}\label{sec:conclusions}

We have complemented the panchromatic study of the 11 DSFGs, spectroscopically confirmed to be at the peak of Cosmic SFH, presented in \citet[][]{Pantoni2021:2021MNRAS.504..928P}, by focusing on the ALMA view of the galaxies. We selected the 11 objects in the (sub-)millimeter regime requiring the following criteria to be fulfilled for each galaxy: 3 or more detections in the optical domain; 6 or more detections in the NIR+MIR bands; 2 or more detections in the FIR band; spectroscopically confirmed redshift in the range $1.5<z<3$; 1 or more detections and/or upper limits in the radio and X-ray regimes.
The sources are located in one of the deepest multi-band field currently available, the GOODS-S. In \citet[][]{Pantoni2021:2021MNRAS.504..928P} we exploited the photometry from the X-ray to the radio band to model galaxy SED and extract the main astrophysical properties of the 11 DSFGs (e.g., SFR, stellar mass, stellar attenuation law by dust, dust temperature, IR luminosity, dust and gas mass, AGN fraction). 

In this work we exploited the most recent ALMA continuum maps and spectroscopic images of our 11 DSFGs, selected to have the highest spatial and spectral resolution between the ones publicly available in the ALMA Archive. We derived the dust emission size (or just an upper limit in case of scarce spatial resolution) of the 11 DSFGs in our sample; we analyzed the CO $J>1$ emission lines that we found in the ALMA archive for four of our galaxies (i.e. UDF1, UDF3, UDF8, ALESS067.1) and derived the molecular hydrogen mass; we compare the outcomes in the (sub-)millimeter regime with galaxy emission in the other spectral bands, such as optical, X-ray and radio. Finally we used our findings to put each galaxy in the broader context of galaxy formation and evolution, mainly by referring to the \textit{in-situ} galaxy-BH co-evolution scenario \citep[e.g.,][]{Mancuso2016a:2016ApJ...823..128M, Mancuso2016b:2016ApJ...833..152M, Mancuso2017:2017ApJ...842...95M,Lapi2018a:2018ApJ...857...22L,Pantoni2019:2019ApJ...880..129P}.
In the following we summarize our main results.
\begin{itemize}
    \item We derived the ALMA continuum size of our 11 DSFGs by using the ALMA map at higher spatial resolution currently available in the ALMA Archive (ALMA B6, B7, B9). We performed a 2D Gaussian fitting of each source on the \textit{science-ready} image using the task \textit{imfit} embedded in the CASA viewer. More than half of the sources (six out of eleven) are resolved and the median physical size is 1.15 kpc. We interpret this radius as a measure of the region where the bulk of dusty star formation is occurring. It results to be very compact and always $<5$ kpc.
    \item We compared the ALMA continuum sizes with the HST/WFC3 H$_{160}$ radii \citep[][]{vanderWel2012:2012ApJS..203...24V}. The latter spectral band samples the optical rest-frame emission from stars, and traces the star formation of the galaxy that is mildly obscured by dust or unobscured. We found a median ratio between HST and ALMA sizes of r$_{\rm H}/$r$_{\rm ALMA}=2.05$ and it is always $>1$, in accordance e.g. with the prediction by the \textit{in-situ} scenario for the evolution of high-z massive star forming galaxies \citep[see][]{Lapi2018a:2018ApJ...857...22L} and with other recent works on high-z DSFGs \citep[e.g.][]{Barro2013:2013ApJ...765..104B, Barro2016I:2016ApJ...820..120B, Barro2016II:2016ApJ...827L..32B, Ikarashi2015:2015ApJ...810..133I, Simpson2015:2015ApJ...807..128S, Talia2018:2018MNRAS.476.3956T, Tadaki2020:2020ApJ...901...74T, Massardi2018:2018A&A...610A..53M}. 
    After correcting for the astrometric shift between H$_{160}$ HST maps and ALMA maps, the bulks of the stellar and dust emission coincide within the uncertainties of the astrometric correction ($\sim100$ mas) and the beam resolution, for most of the galaxies. An eventual remaining shift tells us that the peaks of optical and (sub-)millimeter emission are not exactly coincident, due to dust obscuration of stellar light.
    \item We analysed the five CO $J>1$ emission lines that we found for four of our 11 DSFGs, i.e. UDF1, UDF3, UDF8 and ALESS067.1, and derived the (sub-)millimeter redshift of the sources. The double-peaked spectral line profile of three CO lines, along with their velocity and velocity dispersion maps, are consistent with both a rotating disc of molecular gas and an AGN outflow. In particular, the latter is the case of ALESS067.1, for which we detected two CO lines ($J=3$ and $J=6$). The mom1 and mom2 maps suggest the presence of an outflow, even if the associated velocity that is $<500$ km/s does not allow us to confirm this scenario. 
    ALESS067.1 CO(6-5) line is narrow and could trace also the warm star formation in the galaxy (to confirm this interpretation it is necessary to reconstruct the CO SLED). The narrow single-peaked CO(3-2) line of UDF1 is consistent with a \textit{faced-on} rotating molecular gas disc.
    \item We derive the molecular gas mass of UDF1, UDF3, UDF8 and ALESS067.1 from the lower-J CO emission line luminosity, by assuming the CO excitation ladder by \citet[][]{Daddi2015:2015A&A...577A..46D}. The median molecular hydrogen gas mass is M$_{H_2}\sim 1.4\times10^{11}$ M$_\odot$. We measured the CO emission size from the distance between the two components peaks (in case of a double peaked profile) or we give just an upper limit on it, since the sources are not resolved in the mom0 maps. The median CO size is of 3.5 kpc. The CO emission extends over an area greater or equal to the dust continuum size.
    \item We complemented these results by exploiting multi-wavelength images from public catalogs, that allowed us to include in our final interpretation every signature of galaxy merging/interactions and AGN feedback. The compact FIR and radio sizes ($\lesssim$ a few kpc) of our DSFGs, together with their optical radii ($\sim 2-7$ kpc), suggest that the bulk of their star formation can be traced back to in-situ condensation processes. Almost half of our sources shows an optical isolated morphology, while six galaxies (UDF5, UDF10, UDF11, ALESS067.1, AzTEC.GS21, AzTEC.GS22) have more complex (i.e., clumpy or disturbed) morphologies, but still on scales $\sim5-10$ kpc. They may indicate the presence of minor interactions - that can prolong the star formation in the dominant galaxy by refuelling it with gas - or may be just a signature of the ongoing dusty star formation. 
    \item We can state that most of the galaxies in our sample are caught in the \textit{gas compaction} phase and models predict that they should be quenched by the AGN feedback in $\sim 10^{8}$ yr. Three objects show some features that can be interpreted as signatures of nuclear activity by the detection of possible AGN-driven molecular outflows (UDF3, UDF8, ALESS067.1). After quenching, galaxy evolution should be mainly driven by stellar populations aging and mass additions by dry merger events. Ultimately, we expect our 11 DSFGs to become compact quiescent galaxies or massive ETGs.
    \item We gathered together all the evidences coming from galaxy multi-wavelength emission, gas spectroscopy and imaging at highest resolution currently available, along with a possible self-consistent and physical motivated theoretical model describing massive star forming galaxy evolution \citep[e.g.,][]{Mancuso2016a:2016ApJ...823..128M, Mancuso2016b:2016ApJ...833..152M, Mancuso2017:2017ApJ...842...95M,Lapi2018a:2018ApJ...857...22L,Pantoni2019:2019ApJ...880..129P}. 
    In such a way, we provide a novel approach in characterizing the individual DSFGs and predicting their subsequent evolution.
    \item Finally, we stress the need of more sensitive multi-wavelength maps and higher spatial and spectral resolution images in the diverse spectral bands to definitely clarify the relative impact of in-situ processes, galaxy interactions and AGN feedback in determining massive star forming galaxy evolution at high-z and their morphological transformation.
\end{itemize}

\section*{Acknowledgements}

The authors thank the anonymous referee for stimulating and constructive suggestions that helped to improve this study.
LP gratefully thanks M. Bischetti, S. Campitiello and C. Memo for the useful and helpful discussions, comments and/or support.

This paper makes use of the following ALMA data: ADS/JAO.ALMA\#2012.1.00173.S (PI: Dunlop); \#2012.1.00307.S (PI: Hodge); \#2012.1.00983.S (PI: Leiton); \#2015.1.00098.S (PI: Kohno); \#2015.1.01074.S (PI: Inami); \#2016.1.00564.S (PI: Weiss); \#2016.1.01079.S (PI: Bauer);\#2017.1.00270.S (PI: Walter); \#2017.1.00001.S (PI: Rujopakarn); \#2017.1.01347.S (PI: Pope); \#2019.2.00246.S (PI: Calistro Rivera).
ALMA is a partnership of ESO (representing its member states), NSF (USA), and NINS (Japan), together with NRC (Canada), NSC and ASIAA (Taiwan), and KASI (Republic of Korea), in cooperation with the Republic of Chile. The Joint ALMA Observatory is operated by ESO, AUI/NRAO, and NAOJ.

We acknowledge financial support from the grants: PRIN MIUR 2017 prot. 20173ML3WW 001 and PRIN MIUR 2017 prot. 20173ML3WW 002 (‘Opening the ALMA window on the cosmic evolution of gas, stars and massive black holes’).
A. Lapi is supported by the EU H2020-MSCAITN-2019 Project 860744 ‘BiD4BEST: Big Data applications for Black hole Evolution STudies’. 
\section*{Data Availability}

This article uses public data products from ALMA Archive (repository available at the following link: https://almascience.nrao.edu/asax/). Project codes of interest are listed in the Acknowledgements.

H$_{160}$ (F160W) HST/WFC3 images are taken from the Hubble Legacy Archive (https://hla.stsci.edu/hlaview.html).

Photometry in the optical, infrared, radio and X-rays comes from (in the order): 
\begin{itemize}
    \item GOODS-MUSIC sample: a multi-colour catalog of near-IR selected galaxies in the GOODS-South field \citep[][]{Grazian2006:2006A&A...449..951G}, available at the link: https://cdsarc.unistra.fr/viz-bin/cat/J/A+A/449/951 - VizieR DOI: 10.26093/cds/vizier.34490951;
    \item combined PEP/GOODS-Herschel data of the GOODS fields by \citet[][https://www.mpe.mpg.de/ir/Research/PEP/DR1]{Magnelli2013:2013A&A...553A.132M} and publicly available at \\http://www.mpe.mpg.de/ir/Research/PEP/public\_data\_releases.php \citep[see also][https://cdsarc.unistra.fr/viz-bin/cat/J/A+A/528/A35 - VizieR DOI: 10.26093/cds/vizier.35280035]{Magnelli11:2011yCat..35280035M};
    \item Herschel Multi-tiered Extragalactic Survey: HerMES \citep[][]{Oliver14:2014yCat.8095....0O, Oliver17:2017yCat.8103....0H}, publicly available through the Herschel Database in Marseille, HeDaM, at http://hedam.oamp.fr/HerMES and https://cdsarc.unistra.fr/viz-bin/cat/VIII/95 and https://cdsarc.unistra.fr/viz-bin/cat/VIII/103 (VizieR);
    \item Very Large Array 1.4 GHz survey of the Extended Chandra Deep Field South: second data release \citep[][]{Miller13:2013yCat..22050013M}, https://cdsarc.unistra.fr/viz-bin/cat/J/ApJS/205/13 - VizieR DOI: 10.26093/cds/vizier.22050013;
    \item Very Large Array 6 GHz imaging by \citet{Rujopakarn2016:2016ApJ...833...12R}: Project ID VLA/14A-360 \citep[follow-up of the B6 ALMA sources by][in the HUDF-S; ADS/JAO.ALMA\#2012.1.00173.S]{Dunlop2017:2017MNRAS.466..861D}.
    \item Chandra Deep Field-South survey: 7 Ms source catalogs \citep[][]{Luo17:2017yCat..22280002L}, https://cdsarc.unistra.fr/viz-bin/cat/J/ApJS/228/2 (VizieR).
\end{itemize}



\bibliographystyle{mnras}
\bibliography{main} 




\appendix

\section{Individual source analysis}\label{appendix}

The panchromatic approach already presented in \citet[][]{Pantoni2021:2021MNRAS.504..928P} and here enriched, by the analysis of multi-wavelength spatially-resolved emission and CO spectral lines, allows us to provide a self-consistent characterization of the 11 DSFGs, that includes their role in the framework of galaxy formation and evolution. In this Appendix we put together all the observational evidences and give an insight on the possible evolutionary scenario for the individual galaxies that are in our sample.

While characterizing galaxy cold gas phase we preferentially use the gas mass measurements from CO spectral lines (when available; see Tab. \ref{tab:summary_resolved}) rather than the ones from dust continuum (Tab. \ref{tab:summary_integral}), since the latter can miss the gas distributed over scales larger than the dust continuum size \citep[e.g.,][]{Riechers2011:2011ApJ...739L..31R, Hodge2015:2015ApJ...798L..18H, Spilker2015:2015ApJ...811..124S, Decarli2016II:2016ApJ...833...70D}.

Finally, we note to the reader that all the multi-wavelength radii and sizes presented here are circularized (see Eq. \eqref{eq_sizes}).

\subsection{UDF1 (J033244.01-274635.2)}

UDF1 is a DSFG at $z_{\rm spec}=2.698$ (see Tab. \ref{tab:galaxy_properties}). We measured its spectroscopic redshift from a CO(3-2) line (Tab. \ref{tab:CO_gas_mol}) that is consistent, within the errors, with the redshift of the C[II] emission line at 8600\AA{} (2330\AA{} rest frame) by \citet[][; HII region-type spectrum]{Szokoly2004:2004ApJS..155..271S}. UDF1 was detected in the ALMA B6 ($\lambda\sim1.3$ mm) survey by \citet{Dunlop2017:2017MNRAS.466..861D} and more recently by \citet{Franco2018:2018A&A...620A.152F} at $\lambda\sim1.1$ mm (source ID: AGS6). \citet{Rujopakarn2016:2016ApJ...833...12R} detected its radio counterpart with the VLA at 6 GHz. The corresponding fluxes are S$_{\rm 1.3\,mm}=924\pm76$ $\mu$Jy, S$_{\rm 1.1\,mm}=1.26\pm0.16$ mJy and S$_{\rm 6\,GHz}=9.02\pm0.57$ $\mu$Jy. In the ALMA archival continuum map at the best spatial resolution currently available for UDF1 (Project code: 2017.1.00001.S, ALMA B7; see Tab. \ref{tab:continuum}), the source is detected at a significance level $>5\sigma$. We measure a flux S$_{\rm 1\,mm}=2900\pm300\,\mu$Jy (Tab. \ref{tab:continuum}). This flux is consistent with the 850 $\mu$m ALMA B7 flux measured by \citet{Cowie2018:2018ApJ...865..106C}, i.e. S$_{\rm 850\,\mu m}=3.38\pm0.32$ mJy (source ID: no. 22; name ALMA033244-274635), and by \citet[][]{Rujopakarn2019:2019ApJ...882..107R} on the same ALMA map, i.e. S$_{\rm B7}=3407\pm226$ $\mu$Jy. In the B7 ALMA map the source is resolved: we measure a circularized radius r$_{\rm ALMA}\simeq0.46$ kpc, where we expect the dusty star-formation to be located.
\citet{Rujopakarn2016:2016ApJ...833...12R} find the 6 GHz radio flux to be compatible with the star formation activity of the ALMA source, confirming the outcome we obtained from the analysis of galaxy SED in \citet[][]{Pantoni2021:2021MNRAS.504..928P}. The radio emission, that samples also the mildly obscured and un-obscured star-formation of the galaxy, extends over a larger region: r$_{\rm VLA}\sim2.7$ kpc \citep[FWHM;][]{Rujopakarn2016:2016ApJ...833...12R}. From the CO(3-2) line luminosity we estimate the molecular hydrogen content of the galaxy to be M$_{H_2}=(2.6\pm0.7)\times10^{10}$ M$_\odot$. From its velocity and velocity dispersion maps we claim that the galaxy cold gas component is probably a \textit{faced-on} rotating disk that extends over a radius $\lesssim3$ kpc (see Tab. \ref{tab:COemission}). Its spatial distribution samples the galaxy star formation, both obscured and un-obscured, as well as the aforementioned radio emission, that extends over a region of a similar size.

The peak of dust radiation matches pretty well that of stellar emission (the HST/WFC3 emission map in the filter H$_{160}$, overlapped by ALMA contours, is shown in Fig. \ref{fig:cont}). The optical size \citep[r$_{\rm H}\sim2.6$ kpc;][see Tab. \ref{tab:galaxy_properties}]{Rujopakarn2019:2019ApJ...882..107R} extends over a radius that is more than a factor 5 larger than the ALMA size, reported above and in Tab \ref{tab:continuum}.
These evidences can be easily ascribed to a very rapid condensation and compaction of the gas towards the centre of the galaxy (over a typical timescale of $\tau\sim10^6$ yr), where the bulk of dusty star formation takes place (M$_{\rm dust}\sim6\times10^8$ M$_\odot$, see Tab. \ref{tab:galaxy_properties} and \ref{tab:summary_integral}).

From the analysis presented in \citet[][]{Pantoni2021:2021MNRAS.504..928P}, UDF1 is almost approaching the galaxy main-sequence \citep[see e.g.][]{Speagle2014:2014ApJS..214...15S} at the corresponding redshift (it is a factor $\sim2\times$ above) with a stellar mass M$_\star=(8\pm1)\times10^{10}$ M$_\odot$ and SFR $=352\pm18$ M$_\odot$ yr$^{-1}$. We think that it is probably experiencing the last phase of dusty star formation before the intervention of the AGN feedback. Indeed, even if the optical counterpart has been historically classified as a normal star-forming galaxy (the presence of an AGN cannot be established from the optical data alone), it has a very powerful counterpart in the X-ray, suggesting the presence of an accreting central SMBH. In the $\simeq$ 7 Ms CDFS catalog by \citet{Luo2017:2017ApJS..228....2L}, the X-ray source (IDX 850) is classified as \textit{AGN} (see Tabs. \ref{tab:galaxy_properties} and \ref{tab:summary_integral}). The intrinsic absorption column density\footnote{The column density $N_H$ (reported here, but also for other sources in our sample, in the following Sections) is derived on the basis of flux ratios in the hard and soft X-ray bands, by assuming a classical AGN slope of 1.8.} $N_H \sim 1.2\times10^{22}$ cm$^{-2}$ gives an absorption-corrected intrinsic $2-10$ keV luminosity of $\sim4\times10^{43}$ erg s$^{-1}$ (see Tab. \ref{tab:summary_integral}). \citet{Rujopakarn2016:2016ApJ...833...12R} suggest that the VLA size of UDF1 can be somehow more extended than the ALMA one due to the central AGN, even if the 6 GHz radio flux does not appear to be enhanced, when compared to the one expected from star formation alone. 
Relying on the prescriptions of the \textit{in-situ} scenario for galaxy evolution \citep[see][]{Mancuso2017:2017ApJ...842...95M}, we expect the nucleus radio emission to emerge from the one associated to star formation only at later times.

Gas and dust rich, with a typical depletion timescale $\tau_{\rm depl}\sim75$ Myr, we think that UDF1 very compact and intense star formation must have triggered a rapid growth of the central SMBH, whose power in the X-ray clearly overwhelms the host galaxy one \citep[][]{Pantoni2021:2021MNRAS.504..928P}. We expect this galaxy to overlap the galaxy main-sequence in a few hundreds of Myr and then to be quenched by the central AGN.

\subsection{UDF3 (J033238.53-274634.6)}

UDF3 is a DSFG at $z_{\rm spec}=2.544^{+0.001}_{-0.002}$ (Tab. \ref{tab:galaxy_properties}), that we measured from a CO(3-2) line detected in a ALMA B3 ($\nu_{\rm obs}\simeq 97.6$ GHz; see Tabs. \ref{tab:COemission}).
Our measurement is consistent, within the errors, with the one by \citet{Decarli2016II:2016ApJ...833...70D}, based on the blind detection of three CO transitions \citep[ASPECS 3 mm.1, 1 mm.1, 1 mm.2;][]{Decarli2016I:2016ApJ...833...69D}, i.e. z$_{\rm spec}=2.543$ \citep[see also][]{Tacconi2018:2018ApJ...853..179T}. Another measurement, consistent with ours, comes from the MUSE HUDF project, i.e.  z$_{\rm spec}=2.541$ \citep[][]{Bacon2017:2017A&A...608A...1B}. Finally, \citet{Momcheva2016:2016ApJS..225...27M} report a grism redshift z$_{\rm spec}\sim2.561$, based on the detection of the [O II] line in the 3D-HST data, that is in accordance with the aforementioned spectroscopic determinations of the source redshift to the second decimal digit.

UDF3 was detected both in the B6 ($\sim1.3$ mm) ALMA continuum and 6 GHz VLA maps of the HUDF by \citet{Dunlop2017:2017MNRAS.466..861D} and \citet{Rujopakarn2016:2016ApJ...833...12R}. The corresponding fluxes are S$_{\rm 1.3\,mm}=863\pm84$ $\mu$Jy and S$_{\rm 6\,GHz}=12.06\pm0.55$ $\mu$Jy.
We found a $>5\sigma$ continuum detection for this source in the ALMA B7 (Project code: 2017.1.00001.S, Tab. \ref{tab:continuum}) and we measured a flux S$_{\rm 1\,mm}=1600\pm200\,\mu$Jy. This flux is in accordance with the 850 $\mu$m ALMA flux measured by \citet{Cowie2018:2018ApJ...865..106C}, i.e. S$_{\rm 850\,\mu m}=2000\pm400\,\mu$Jy (source ID: no. 48; name ALMA033238-274634), and the 1.1 mm ALMA flux measured by \citet{Franco2018:2018A&A...620A.152F}, i.e. S$_{\rm 1.1\,\mu m}=1130\pm105\,\mu$Jy (source ID: AGS12).
\citet{Rujopakarn2016:2016ApJ...833...12R} found the 6 GHz radio flux of UDF 3 to be solely compatible with the star formation activity of the galaxy, in agreement with the outcome from our SED analysis \citep[][]{Pantoni2021:2021MNRAS.504..928P}. However, they claim that the possible presence of a central AGN, could have an impact on the radio size, i.e. r$_{\rm VLA}= 1.5\pm0.1$ kpc \citep[FWHM;][]{Rujopakarn2016:2016ApJ...833...12R}. The ALMA (and VLA) source has an X-ray counterpart (IDX 718) in the $\simeq$ 7 Ms CDFS survey catalog by \citet{Luo2017:2017ApJS..228....2L}. The source $2-10$ keV luminosity is $1.8\times10^{42}$ erg s$^{-1}$. All these evidences suggest that the central BH could be accreting material in the host galaxy nucleus, even if its emission do not clearly overwhelms the X-ray light coming from star formation at the moment \citep[][]{Pantoni2021:2021MNRAS.504..928P}. 

The rest-frame UV/optical morphology is unclear and may be clumpy/disturbed (see Fig. \ref{fig:cont} and Tab. \ref{tab:summary_resolved}). However, we do not find any clear evidence of interactions. The starlight emission is compact \citep[r$_{\rm H}\sim1.6$ kpc;][see Tab. \ref{tab:galaxy_properties}]{vanderWel2012:2012ApJS..203...24V} and its peak appears to be shifted when compared to the ALMA B7 continuum emission, possibly due to dust obscuration of stellar light. (Fig. \ref{fig:cont}). The absence of interactions is confirmed by the velocity and dispersion velocity maps of CO(3-2) molecular line, that are smooth (i.e., not disturbed) and consistent with a rotating disc of molecular gas (confirmed by the double-peaked profile of the spectral line; see Fig. \ref{fig:panel_CO_3sources}) extending over a radius of $\sim1.4$ kpc. The (sub-)millimeter source is resolved in the ALMA B7 continuum map: we measure a circularized radius of r$_{\rm ALMA} \sim 0.6$ kpc (Tab. \ref{tab:continuum}). However, to confirm this scenario we need multi-wavelength images at higher spatial resolution.

In the analysis presented in \citet[][]{Pantoni2021:2021MNRAS.504..928P}, we found UDF3 to be a very young object (age $\sim200-300$ Myr), forming stars at a rate SFR $\sim500$ M$_\odot$ yr$^{-1}$ (sSFR $\sim5.8$ Gyr$^{-1}$). Fuelled by the large amount of gas and dust (M$_{H_2}=(1.5\pm0.3)\times10^{11}$ M$_\odot$ and M$_{\rm dust}\sim4\times10^8$ M$_\odot$; Tabs. \ref{tab:CO_gas_mol} and \ref{tab:galaxy_properties}), its intense burst of star formation (depletion timescale $\tau_{\rm depl}\lesssim100$ Myr) must trigger the growth of the central BH, that we expect to become active and quench star formation in a few $10^8$ yr. In the SFR$-{\rm M}_{\star}$ plot, UDF3 lies to the top-left side of the main-sequence at the corresponding redshift \citep[even if it is still consistent with the $2\sigma$ scatter of the relation; see][]{Speagle2014:2014ApJS..214...15S, Pantoni2021:2021MNRAS.504..928P}. A similar result was found by \citet{Elbaz2018:2018A&A...616A.110E}, who classify UDF3 as a starburst galaxy.

\subsection{UDF5 (J033236.94-274726.8)}

UDF5 is a DSFG detected both in the B6 ($\sim1.3$ mm) ALMA and 6 GHz VLA maps of the HUDF by \citet{Dunlop2017:2017MNRAS.466..861D} and \citet{Rujopakarn2016:2016ApJ...833...12R}. The corresponding fluxes are S$_{\rm 1.3\,mm}=311\pm49$ $\mu$Jy and S$_{\rm 6\,GHz}=6.26\pm0.46$ $\mu$Jy. 

As to the source redshift, \citet{Momcheva2016:2016ApJS..225...27M}  report a $z_{\rm spec}=1.759$, basing on 3D-HST spectroscopy (Tab. \ref{tab:galaxy_properties}). 

For this millimeter source we do not find any X-ray counterpart  in the literature, neither by the association with the $\simeq$ 7 Ms X-ray catalogue by \citet{Luo2017:2017ApJS..228....2L} or in the supplementary catalog at very low significance. This may indicate either that no (i.e. not very powerful) AGN is present or that it is highly obscured (i.e. Compton-thick, with $N_{\rm H}\gtrsim10^{24}$ cm$^{-2}$). Since the source lies in a very deep region of the \textit{Chandra} map (equivalent exposure time of about 6.22 Ms), we think that the most probable hypothesis to explain the non-detection is the latter. Following the \textit{in-situ} scenario predictions, the central BH must have just started to accrete material from the surroundings; as such, we do not expect to observe any signature of its activity on the source morphology. Both ALMA and VLA sizes can provide a good measure of the region where the bulk of star formation is occurring, i.e. very compact, $\sim2$ kpc \citep[r$_{\rm ALMA}<2.5$ kpc; r${\rm VLA}\sim1.5$ kpc; see Tab. \ref{tab:continuum} and][]{Rujopakarn2016:2016ApJ...833...12R}. \citet{Rujopakarn2016:2016ApJ...833...12R} claim that the 6 GHz VLA flux is solely compatible with the star formation activity of the galaxy, confirming the result by \citet[][]{Pantoni2021:2021MNRAS.504..928P}.

 The HST/WFC3 H$_{160}$ circularized radius of UDF5 is r$_{\rm H}\sim2.3$ kpc, slightly larger than the radio one and consistent with the upper limit on the ALMA B6 size. The rest-frame optical morphology is clumpy/disturbed (Fig. \ref{fig:cont} and Tab. \ref{tab:summary_resolved}). Since no clear evidence of interactions and/or AGN feedback has been observed, we ascribe the galaxy multi-band morphology to both the dust enshrouded star formation occurring at radii $\lesssim2$ kpc and the gas compaction towards the central region.

From the analysis by \citet[][]{Pantoni2021:2021MNRAS.504..928P}, UDF5 is in the vicinity of the main-sequence at the corresponding redshift. This young galaxy (age $\sim400$ Myr) shows a more moderate SFR ($\sim 80-90$ M$_\odot$ yr$^{-1}$) and a more modest stellar mass (a few $10^{10}$ M$_\odot$) compared to the other galaxies in our sample (cf. Tab. \ref{tab:galaxy_properties}). The depletion timescale is $\tau_{\rm depl}\lesssim900$ Myr and the dust content is quite high (M$_{\rm dust}\sim4\times10^{10}$ M$_\odot$). Following the \textit{in-situ} scenario, we expect the star-formation to last longer than what is seen for the bulk of high-z DSFGs ($\tau_{\rm burst}\sim$ a few $10^9$ yr). This picture can explain why we do not have any evidence of an active nucleus: less intense burst of star formation implies a slower BH Accretion History and at the current galaxy age (i.e., $\sim400$ Myr) the central BH is highly obscured in the X-ray and the emission is still dominated by star formation.

\subsection{UDF8 (J033239.74-274611.4)}

UDF8 is a DSFG at z$_{\rm spec}\sim1.6$. The slit redshift z$_{\rm spec}=1.552$ by \citet{Kurk2013:2013A&A...549A..63K} matches the CO line detection by \citet[][source ID: ASPECS3mm.2]{Decarli2016II:2016ApJ...833...70D} and our measurement from the CO(2-1) spectral line detected in the ALMA B3 data cube, i.e. $z_{\rm spec}=1.5510^{+0.0014}_{-0.0005}$ (Tab \ref{tab:galaxy_properties}), that we use in our analysis.

UDF8 was detected both in the B6 ($\sim1.3$ mm) ALMA and 6 GHz VLA maps of the HUDF by \citet{Dunlop2017:2017MNRAS.466..861D} and \citet{Rujopakarn2016:2016ApJ...833...12R}. The corresponding fluxes are S$_{\rm B6}=208\pm46$ $\mu$Jy and S$_{\rm 6\,GHz}=7.21\pm0.47$ $\mu$Jy. The source is resolved both in ALMA and VLA maps and the corresponding radii are r$_{\rm ALMA}\sim4.1$ kpc (Tab. \ref{tab:continuum}) and r$_{\rm VLA}\sim2.1$ kpc. This dusty galaxy, still compact, is however more extended in the millimeter than the bulk of high-z DSFGs (typical ALMA radius is found to be $\lesssim1-2$ kpc).

In the rest-frame UV/optical the galaxy appears to be isolated  (see Fig. \ref{fig:cont} and Tab. \ref{tab:summary_resolved}), suggesting that the bulk of star formation can be traced back to local, \textit{in-situ} condensation processes. The galaxy is more extended in the optical (r$_{\rm H}\sim5.7$ kpc) than in the millimeter continuum, consistently with what we expect in the gas compaction evolutionary phase. This is confirmed by the double-peaked spectral profile of its CO(2-1) line and the velocity and velocity dispersion maps (Fig. \ref{fig:panel_CO_3sources}), that suggest we are observing a rotating disc of molecular gas extending over an area of radius $r_{\rm, CO}\sim3.5$ kpc. From the CO line luminosity we derive a molecular hydrogen mass M$_{H_2}=(5.8\pm1.1)\times10^{10}$ M$_\odot$ that is consistent within the uncertainties with the one estimated by \citet{Decarli2016II:2016ApJ...833...70D}, i.e. M$_{H_2}\sim6.5\times10^{10}$ M$_\odot$, under the assumption of the same $\alpha_{\rm CO}=3.6$ K km pc$^2$ s$^{-1}$ M$_\odot^{-1}$.

UDF8 is detected with \textit{Chandra}. We found its X-ray counterpart (IDX 748) in the $\simeq$ 7 Ms CDFS survey catalogue by \citet{Luo2017:2017ApJS..228....2L}. The source $2-10$ keV luminosity is $3.5\times10^{43}$ erg s$^{-1}$, suggesting that UDF8 hosts an active galactic nucleus (see Tab. \ref{tab:galaxy_properties}), which dominates over the X-ray emission of the host galaxy \citep[][]{Pantoni2021:2021MNRAS.504..928P}. 
Even if the galaxy host an X-ray AGN, \citet{Rujopakarn2016:2016ApJ...833...12R} claim that its contribution to the radio emission is negligible. Indeed, they find the radio flux at 6 GHz to be consistent with the star formation alone, as to the radio morphology. From the analysis of galaxy SED by \citet[][]{Pantoni2021:2021MNRAS.504..928P} we obtain the same outcome.

UDF8 is a main-sequence galaxy of age $\sim1$ Gyr, forming stars at a rate SFR $\sim100$ M$_\odot$ yr$^{-1}$, with a sSFR $\sim1.5$ Gyr$^{-1}$ \citep[][]{Pantoni2021:2021MNRAS.504..928P}. We expect the AGN quenching to be very close: eventually, its effect could be  already seen under the shape of outflows and winds that can affect the multi-band sizes of UDF8 and broaden its CO(2-1) spectral line profile (with respect to pure disc rotation). To confirm this scenario we need multi-band images at higher spatial resolution of the galaxy nuclear region.

\subsection{UDF10 (J033240.73-274749.4)}

UDF10 is a DSFG at z$_{\rm spec}=2.086$ \citep[grism redshift by][based on the detection of an optical rest-frame line in the 3D-HST data; Tab. \ref{tab:galaxy_properties}]{Momcheva2016:2016ApJS..225...27M}. 

The galaxy was detected at 3.6$\sigma$ in the ALMA B6 (S$_{\rm B6}=184\pm46\,\mu$m) in the blind-survey of the HUDF by \citet{Dunlop2017:2017MNRAS.466..861D}. In the radio \citep[6 GHz VLA][]{Rujopakarn2016:2016ApJ...833...12R} it is detected at a  significance level $<3\sigma$, thus only an upper limit of the 6 GHz radio flux of the source is available, i.e. S$_{\rm 6\,GHz}\lesssim 0.70$ $\mu$Jy.
The source is not resolved in the ALMA map, but its size must be $<2.5$ kpc (Tab. \ref{tab:continuum}). After the astrometric corrections, the ALMA centroid emission appears to be shifted of $\sim10$ mas from the optical peak \citep[Fig. \ref{fig:cont}; see also][]{Dunlop2017:2017MNRAS.466..861D}. 
UDF10 optical radius is r$_{\rm H}\sim2.0$ kpc (Tab. \ref{tab:galaxy_properties}), that is comparable with the upper limit on ALMA size.

UDF10 is detected with \textit{Chandra} (IDX 756) in the $\simeq$ 7 Ms CDFS catalog by \citet{Luo2017:2017ApJS..228....2L}. The source $2-10$ keV luminosity is then $6\times10^{41}$ erg s$^{-1}$. \citet{Luo2017:2017ApJS..228....2L} classified the source to be a normal galaxy (Tab. \ref{tab:galaxy_properties}).

Putting together these evidences with the outcomes from the galaxy SED analysis presented in \citet[][]{Pantoni2021:2021MNRAS.504..928P}, UDF10 is a $z\sim2$ main-sequence galaxy, since it overlaps the corresponding relation by \citet[][]{Speagle2014:2014ApJS..214...15S}. Quite old (age $\sim 1$ Gyr) when compared to the other sources of the sample, it is characterized by a less intense burst of star formation, with SFR $\sim40$ M$_\odot$ yr$^{-1}$ and sSFR is $\sim1-2$ Gyr$^{-1}$. With a slightly smaller dust content (M$_{\rm dust}\sim2\times10^8$ M$_\odot$) than the majority of DSFGs, possibly its star formation has not triggered yet the activity of the central nucleus, whose power is still too low to have any kind of impact on the host galaxy. Indeed, the emission coming from star formation dominates both the radio and the X-rays \citep[][]{Rujopakarn2016:2016ApJ...833...12R, Pantoni2021:2021MNRAS.504..928P}. However, multi-wavelength images at higher spatial resolution and sensitivity are crucial to confirm (or reject) this scenario. 
 
\subsection{UDF11 (J033240.06-274755.5)}

UDF11 is a DSFG at $z_{\rm spec}\sim2$. The most recent redshift measurement comes from the MUSE HUDF project, i.e. z$_{\rm spec} = 1.9962\pm0.0014$ \citep[][]{Bacon2017:2017A&A...608A...1B, Dunlop2017:2017MNRAS.466..861D}, in total agreement with the previous deep-spectroscopy by \citet[][]{Kurk2013:2013A&A...549A..63K}, who exploited the red-sensitive optical spectrograph FORS2 installed at the Very Large Telescope. 

UDF11 was detected both in the B6 ($\sim1.3$ mm) ALMA and 6 GHz VLA maps of the HUDF by \citet{Dunlop2017:2017MNRAS.466..861D} and \citet{Rujopakarn2016:2016ApJ...833...12R}. The corresponding fluxes are S$_{\rm 1.3\,mm}=186\pm46$ $\mu$Jy and S$_{\rm 6\,GHz}=9.34\pm0.74$ $\mu$Jy.

The galaxy is resolved both in ALMA and VLA maps. In the millimeter it results to be more extended (r$_{\rm ALMA}\sim3.4$ kpc; see Tab. \ref{tab:continuum}) than the bulk of high-z DSFGs (typical ALMA radius is found to be $\lesssim1-2$ kpc). 
The radio morphology is well reproduced by a two-components fit: one is spatially coincident with the central ALMA emission (r$_{\rm VLA}\sim3.4$ kpc); the other, more compact (r$_{\rm VLA}\sim0.7$ kpc) but shifted to the sides of the central millimeter emission, could indicate the presence of two radio lobes/hot spots, suggesting the presence of a central radio AGN \citep[][]{Rujopakarn2016:2016ApJ...833...12R}. The radio flux at 6 GHz is solely consistent with the ongoing star formation in the host galaxy, as it is confirmed by the SED analysis presented in \citet[][]{Pantoni2021:2021MNRAS.504..928P}. 
Basing on these results we can assert that the central nucleus is experiencing the Radio Quiet (RQ) phase, producing jets that are still not dominant in the radio band. This is partially confirmed by the properties of the X-ray source counterpart (IDX 751) in the \textit{Chandra} 7Ms catalog by \citet{Luo2017:2017ApJS..228....2L}. The intrinsic absorption column density $N_H \sim 3.79\times10^{21}$ cm$^{-2}$ gives an absorption-corrected intrinsic $2-10$ keV luminosity of $1.7\times10^{42}$ erg s$^{-1}$. \citet{Luo2017:2017ApJS..228....2L} do not find any clear evidence to classify the source as \textit{AGN}
(see Tab. \ref{tab:galaxy_properties}).  However, an X-ray luminosity $>10^{42}$ erg s$^{-1}$ could indicate that a (small) fraction of the emission may be traced back to the central engine. We expect the galaxy to host a RQ AGN, still accreting material.

Also the HST/WFC3 H$_{160}$ radius is more extended (r$_{\rm H}\sim4.5$ kpc; see Tab. \ref{tab:galaxy_properties}) than the aforementioned ALMA and VLA sizes. The rest-frame optical morphology is clearly clumpy/disturbed (Fig. \ref{fig:cont}). This evidence could be traced back to some interactions with the ambient, replenishing with gas the galaxy at large radii and fuelling the star formation (that is still ongoing, even if a signature of AGN feedback is observed in the radio band). Alternatively, the optical clumpy morphology could be simply ascribed to the combination of quenching and host galaxy star formation. Note that AGN driven winds and outflows can locally have a positive impact on star formation, compressing the gas phase and increasing its density at large radii, i.e. r $\gtrsim1$ kpc \citep[e.g.,][]{CresciMaiolino2018:2018NatAs...2..179C, Shin2019:2019ApJ...881..147S}. This may have affected the SED derived age of the galaxy (which appears younger) and could justify the still high SFR. Multi-wavelength imaging at higher resolution are crucial to shed light on this respect.

From the SED analysis by \citet[][]{Pantoni2021:2021MNRAS.504..928P}, the ongoing star formation burst in UDF11 has an age $\sim400$ Myr and it is forming stars at a rate SFR $\sim250$ M$_\odot$ yr$^{-1}$, with a sSFR $\sim4$ Gyr$^{-1}$. Showing a more modest gas and dust content (i.e., M$_{\rm gas}\sim6\times10^{9}$ M$_\odot$ and M$_{\rm dust}\sim(1.5\times10^8$ M$_\odot$) than the majority of our DSFGs, UDF11 is characterized by a depletion timescale of $\tau_{\rm depl}\sim25$ Myr. We expect the AGN quenching to be close, as it is also confirmed by the extended radio size of the galaxy and the disturbed morphology in the optical, that may be affected by the energetic of the central nucleus. Indeed, in the SFR$-{\rm M}_\star$ plot, UDF11 is indeed almost on the main-sequence at the corresponding redshift.

\subsection{UDF13 (J033235.07-274647.6)}

UDF13 is a DSFG at redshift z$_{\rm spec}=2.497\pm0.008$ \cite[grism redshift by][based on the 3D-HST spectroscopy; see Tab. \ref{tab:galaxy_properties}]{Momcheva2016:2016ApJS..225...27M}. 

UDF13 was detected both in the B6 ($\sim1.3$ mm) ALMA and 6 GHz VLA maps of the HUDF by \citet{Dunlop2017:2017MNRAS.466..861D} and \citet{Rujopakarn2016:2016ApJ...833...12R}. The corresponding fluxes are S$_{\rm 1.3\,mm}=174\pm45$ $\mu$Jy and S$_{\rm 6\,GHz}=4.67\pm0.53$ $\mu$Jy. We measure an ALMA B7 continuum flux of S$_{\rm B7}=910\pm170\,\mu$Jy and give an upper limit on the ALMA size, i.e. $<0.65$ kpc (Tab. \ref{tab:continuum}). We note that the optical radius (r$_{\rm H}\sim1.2$ kpc; Tab. \ref{tab:galaxy_properties}) is more extended than the upper limit on the millimeter size. In the UV/optical rest-frame (Fig. \ref{fig:cont}) UDF13 appears as an isolated object with a smooth (i.e. undisturbed) morphology.

UDF13 has a X-ray counterpart (IDX 655) in the 7 Ms \textit{Chandra} catalog by \citet{Luo2017:2017ApJS..228....2L} with a $2-10$ keV luminosity of $1.3\times10^{42}$ erg s$^{-1}$. Luo et al. classify it as \textit{AGN} (Tab. \ref{tab:galaxy_properties}). 
From the mutual analysis of FIR and radio fluxes of the source, \citet{Rujopakarn2016:2016ApJ...833...12R} found the radio emission to be enhanced by the central AGN detected in the X-ray. This conclusion is derived comparing the observed $S_{5\,\rm cm}/S_{1.3\,\rm mm}$ flux ratio with the one predicted by the \citet{Rieke2009:2009ApJ...692..556R} IR SED libraries, that are calibrated on local ULIRGs. This calibration on local Universe dusty galaxies along with the SED libraries intrinsic uncertainties might alter significantly the analysis. From the SED fitting and radio analysis presented in \citet[][]{Pantoni2021:2021MNRAS.504..928P}, the VLA flux is instead consistent with the ongoing star formation in the host galaxy.

The analysis presented in \citet[][]{Pantoni2021:2021MNRAS.504..928P} classifies UDF13 a main-sequence galaxy, confirming the outcome by \citet{Elbaz2018:2018A&A...616A.110E}.
Quite old (age $\sim 900$ Myr) when compared to the other sources of the sample, UDF13 forms stars at a rate SFR $\sim100$ M$_\odot$ yr$^{-1}$, with a sSFR $\sim1.7$ Gyr$^{-1}$. Showing a more modest modest gas and dust content (M$_{\rm gas}\sim5\times10^{9}$ M$_\odot$ and M$_{\rm dust}\sim1.2\times10^8$ M$_\odot$) than the majority of our DSFGs, UDF13 is characterized by a depletion timescale of $\tau_{\rm depl}\sim45$ Myr. We expect this galaxy to be soon quenched by the central AGN and to subsequently become a red and dead galaxy.

\subsection{ALESS067.1 (J033243.19-275514.3)}\label{app:ALESS067.1}

ALESS067.1 is a DSFG at redshift z$_{\rm spec}=2.1212^{+0.0014}_{-0.0005}$ (this work from CO(3-2) spectral line, see Tab. \ref{tab:galaxy_properties}). It is consistent with the one measured from a H$_\alpha$ line with the Gemini Near-Infrared Spectrograph (GNIRS) by \citep[][]{Kriek2007:2007ApJ...669..776K}, i.e. z$_{\rm spec}=2.122$. 

The galaxy (sub-)millimeter counterpart was firstly observed by LABOCA \citep[LESS;][]{Weiss2009:2009ApJ...707.1201W} and then by ALMA as a part of the ALESS project \citep[][]{Smail&Fabin2014:2014Msngr.157...41S}. It is a compact millimeter source, characterized by a radius of r$_{\rm ALMA}\sim1.1$ kpc \citep[Tab. \ref{fig:cont}; but see also][]{Thomson2014:2014MNRAS.442..577T, Fujimoto2017:2017ApJ...850...83F}.
We found two continuum detection in ALMA B3 and B4 (Project Code: 2016.1.00564.S and 2015.1.00948.S) and we measure the fluxes S$_{\rm B3}=60\pm20$ $\mu$Jy and S$_{\rm B4}=190\pm70$ $\mu$Jy.

\begin{figure}
    \centering
    \includegraphics[width=\columnwidth]{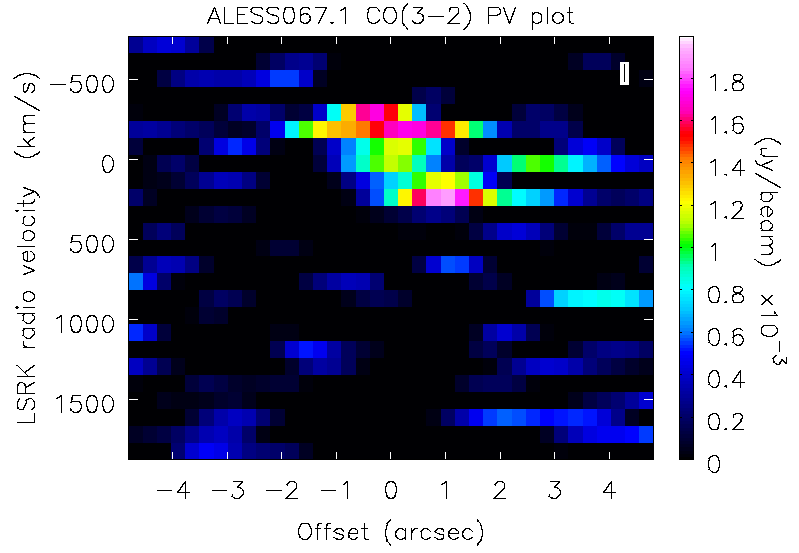}
    \caption{Position - Velocity diagram for ALESS067.1 CO(3-2) molecular emission line.}
    \label{fig:pv_plot}
\end{figure}

In the HST/WFC3 H$_{160}$ image (Fig. \ref{fig:cont}) the galaxy shows an extended (i.e. r$_{\rm H}\sim6.5$ kpc; Tab. \ref{tab:galaxy_properties}) and disturbed/clumpy morphology \citep[][]{Targett2013:2013MNRAS.432.2012T,vanderWel2012:2012ApJS..203...24V}, that could be interpreted either as a signature of an ongoing heavily obscured and intense star formation episode, or as a possible indication of some companions \citep[][claim it is the dominant galaxy of a multiple system]{Targett2013:2013MNRAS.432.2012T}. We can gain some indications on the overall picture by analysing the two CO $J>0$ emission lines detected for ALESS067.1 (see Tab. \ref{tab:COemission} and Fig. \ref{fig:panel_CO_ALESS0671}).
The velocity map (mom1) of the CO(3-2) line show a clear velocity gradient that could indicate the presence of a rotating disc of molecular gas. However, the velocity dispersion map (mom2) shows an evident central peak with a high-velocity tail towards the upper left corner, that could indicate the presence of an outflow, possibly due to a central AGN. Similarly, the double-peaked line profile (Fig. \ref{fig:panel_CO_ALESS0671}) could be ascribed either to the inclination of the rotation plane with respect to the line of sight or to the presence of a double outflow powered by the active nucleus, where one jet is receding and the other is approaching the observer. Other hints can be gained by the analysis of the position-velocity (pv) diagram (Fig. \ref{fig:pv_plot}); reference axis go through the centre and follows the velocity gradient in the velocity map with an inclination of $\sim70$ degrees). The asymmetric and disturbed appearance of the pv plot may confirm the presence of a double AGN-driven molecular outflow, with a  $|v|\simeq400$ km s$^{-1}$ and a peak of $v\simeq900$ km s$^{-1}$ (light cyan structure, that actually, might be just noise). A clear interpretation requires higher spectral and spatial resolution in the ALMA cube, that would have allowed to model the CO(3-2) kinematics, exploiting e.g., 3D-Barolo \citep[][]{DiTeodoro&Fraternali2015:2015MNRAS.451.3021D}, a tool for fitting 3D tilted-ring models to emission-line data cubes that takes into account the effect of beam smearing. 
However, we note that the presence of a molecular outflow triggered by a central engine is consistent also with the CO(6-5) velocity map (mom1; Fig. \ref{fig:panel_CO_ALESS0671}), that shows a peak at higher velocities (i.e. $\sim 100$ km/s) in the orthogonal direction with respect the CO(3-2) peak in the mom1 map.

Nevertheless, ALESS067.1 was found to host a central obscured X-ray AGN with intrinsic L$_{2-10\,{\rm keV}}\simeq3.8\times10^{42}$ erg s$^{-1}$, corrected for obscuration \citep[][]{Luo2017:2017ApJS..228....2L}, whose activity could have an impact on the unclear galaxy morphology. The nucleus emission does not emerge significantly neither in the radio or IR domains \citep[][]{Thomson2014:2014MNRAS.442..577T}.

\begin{figure}
	\includegraphics[width=\columnwidth]{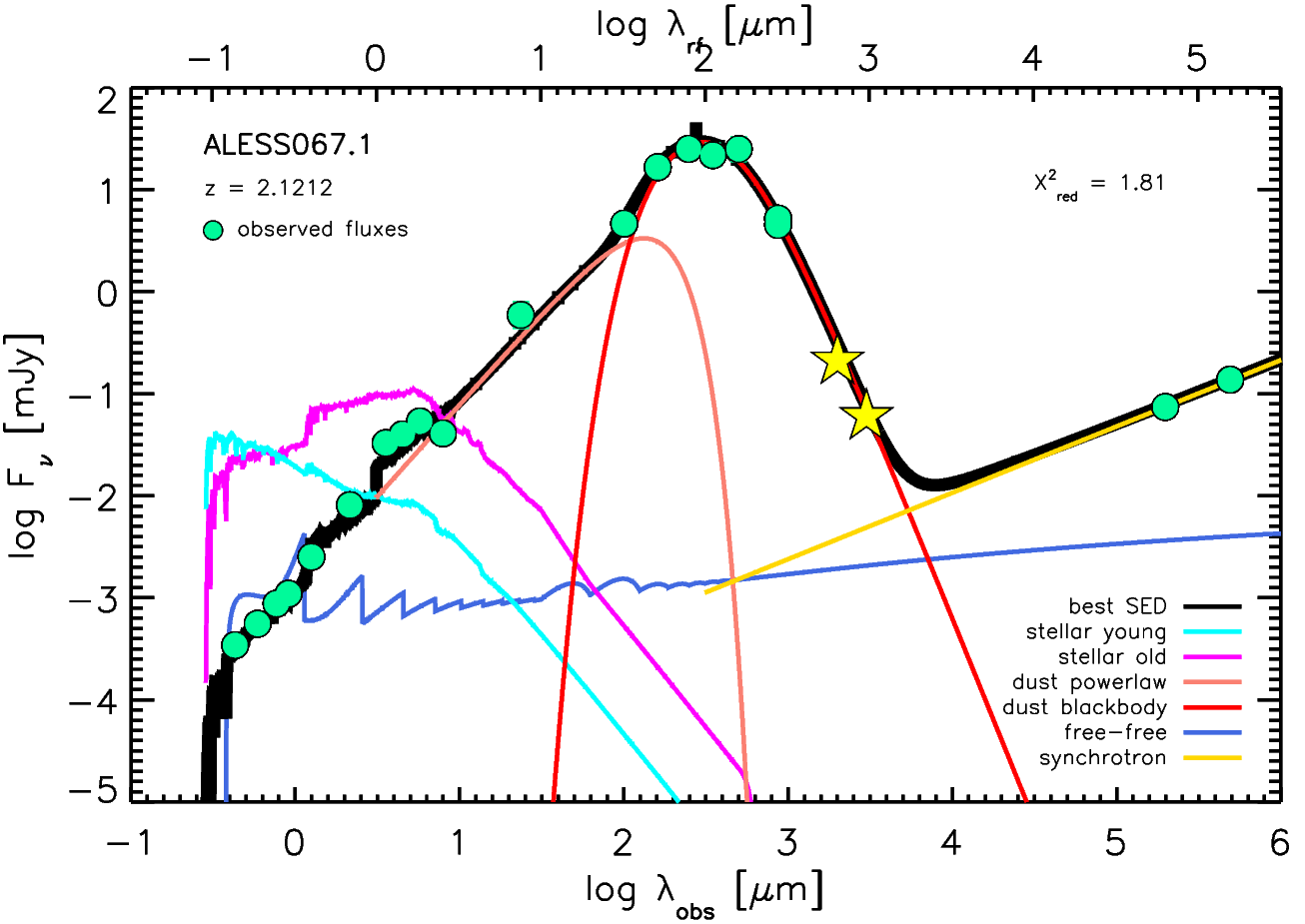}
    \caption{UV-radio best SED for ALESS067.1. Yellow filled stars stand for the ALMA B4 and B3 fluxes measured in this work (see B3 continuum contours in Fig. \ref{fig:panel_CO_ALESS0671}) and are not included in the SED fitting. Green filled circles represent the multi-band available photometry. Error bars are omitted when comparable to symbol size.}
    \label{fig:ALESS067.1_SED}
\end{figure}

In Fig. \ref{fig:ALESS067.1_SED} we show the full SED of ALESS067.1, already studied in \citet[]{Pantoni2021:2021MNRAS.504..928P}. Here we complete the modelling by including the two galaxy radio fluxes available for the source \citep[cf. their Tab. 2][]{Pantoni2021:2021MNRAS.504..928P} in the SED fitting. CIGALE modules in the radio band include synchrotron and free-free emission. The resulting stellar mass M$_\star=(2.9\pm0.3)\times10^{11}$ M$_\odot$ and star formation rate SFR $=485\pm24$ M$_\odot$ yr$^{-1}$ place the source on the main-sequence of star-forming galaxies at the corresponding redshift \citep[e.g.][]{Speagle2014:2014ApJS..214...15S}. An almost constant SFH \citep[e.g.][]{Mancuso2016b:2016ApJ...833..152M} leads to a burst age of $\simeq0.9$ Gyr. The quite huge dust content (M$_{\rm dust}=(5\pm2)\times10^8$ M$_\odot$) is consistent both with predictions from theory \citep[e.g.][]{Popping2017:2017MNRAS.471.3152P, Pantoni2019:2019ApJ...880..129P} and measurements on statistical sample of DSFGs \citep[e.g.][]{Magdis2012;2012ApJ...760....6M}. We find a radio spectral index $\alpha=0.7\pm0.1$ and a FIR/radio flux ratio q$_{\rm IR}=2.5\pm0.1$, in total agreement with the findings by \citet[]{Ibar2009:2009MNRAS.397..281I, Ibar2010:2010MNRAS.401L..53I} and \citet[]{Thomson2014:2014MNRAS.442..577T} for statistical samples of SMGs.

We derive ALESS067.1 molecular hydrogen mass from the CO(3-2) line luminosity as described in Sect. \ref{sec:gasmass}, i.e. M$_{\rm H_2}=(1.68\pm0.5)\times10^{11}$ M$_\odot$, while its dust mass is M$_{\rm dust}\sim1\times10^9$ M$_\odot$ (Tab. \ref{tab:galaxy_properties}. The depletion timescale is $\tau_{\rm depl}\sim 350$ Myr, in agreement with typical values found in literature for high-z DSFGs \citet[e.g.]{CaseyNarayananCooray2014:2014PhR...541...45C}.

Multiple evidences (such as its X-ray $2-10$ keV luminosity, that is $>10^{42}$ erg s$^{-1}$, and the eventual molecular outflow observed in the B3 ALMA map, mentioned before) suggest a forthcoming quenching of star formation in the host galaxy by the AGN.

In order to surely identify (eventual) interactions with minor companions and/or the surrounding ambient and to understand their role and the role of the active nucleus in shaping ALESS067.1 evolution, we need multi-wavelength images at higher sensitivity and spatial/spectral resolution. However, even if we accept the scenario in which the clumpy morphology of the galaxy in the optical is an evidence of an interacting multiple system, we do not expect the presented picture to be significantly altered, since ALESS067.1 is thought to be by far the dominant object \citep[][]{Targett2013:2013MNRAS.432.2012T}.

\subsection{AzTEC.GS25 (J033246.83-275120.9)}

AzTEC.GS25 is a DSFG galaxy firstly detected at 1.1 mm with AzTEC/ASTE (S$_{\rm 1.1\,mm}=1.9\pm0.6$ mJy). For this source, \citet[][]{Popesso2009:2009A&A...494..443P} report a grism redshift z$_{\rm spec}= 2.292\pm0.001$ (Tab. \ref{tab:galaxy_properties}), based on the detection of rest-frame optical line by the VLT/VIMOS spectroscopy. 

We found an ALMA counterpart of the AzTEC source in the DANCING ALMA catalogue by \citet[][source ID: 661]{Fujimoto2017:2017ApJ...850...83F}. The measured a total flux in the ALMA B7 of S$_{\rm 1\,mm}=5.9\pm0.5$ mJy (Tab. \ref{tab:continuum}), in total accordance with the ALMA Band 7 flux by \citet{Cowie2018:2018ApJ...865..106C}, i.e. S$_{\rm 850\,\mu m}=5.9\pm0.18$ mJy (source ID: no. 6; name ALMA033246-275120). Both the millimeter and optical counterparts are resolved and compact. The ALMA radius is r$_{\rm ALMA}\sim1.2$ kpc, while the optical radius is slightly larger, i.e.  r$_{\rm H}\sim1.8$ kpc (tabs. \ref{tab:galaxy_properties} and \ref{tab:continuum}). The smooth morphology of ALMA continuum and optical emission of the galaxy suggest that this object must be isolated. These evidences support the scenario in which the bulk of star formation can be traced back to local, \textit{in-situ} condensation processes.

AzTEC.GS25 has a X-ray counterpart (IDX 844) in the $\simeq 7$ Ms \textit{Chandra} catalog by \citet[][]{Luo2017:2017ApJS..228....2L}. The intrinsic column density $N_H \sim 8.2\times10^{22}$ cm$^{-2}$ gives an absorption-corrected intrinsic $2-10$ keV luminosity of $6\times10^{42}$ erg s$^{-1}$. From the analysis of its X-ray emission, \citet[][]{Luo2017:2017ApJS..228....2L} classified the source as \textit{AGN} (Tab. \ref{tab:galaxy_properties}).

\citet{Yun2012:2012MNRAS.420..957Y} found a radio (unresolved) counterpart of the AzTEC source in the deep ($\sigma\sim8$ $\mu$Jy) VLA 1.4 GHz imaging survey by \citet{Kellermann2008:2008ApJS..179...71K} and \citet{Miller2013:2013ApJS..205...13M}. The corresponding radio flux is S$_{1.4\,GHz}=89.5\pm6.2$ $\mu$Jy, and it is consistent with the host galaxy star formation activity. The red IRAC/MIPS source associated with radio emission GS25a is located only 6.8 arcsec away from the AzTEC centroid position and \citet{Magnelli2013:2013A&A...553A.132M} assumed it as the NIR counterpart of the millimeter source.
The 870 $\mu$m LABOCA source LESS J033246.7-275120 (S$_870\,\mu$m $= 5.9 \pm$ 0.5 $\mu$Jy) is well centred on GS25a, and \citet[][]{Biggs2011:2011MNRAS.413.2314B} also identify the same galaxy as the robust LABOCA counterpart. In \citet[][]{Pantoni2021:2021MNRAS.504..928P}, we used the flux coming from these multi-wavelength counterparts to build the galaxy SED. From the SED analysis, we found AzTEC.GS25 to be a young object (age $\sim300$ Myr), forming stars at a rate SFR $\sim400$ M$_\odot$ yr$^{-1}$, with a sSFR $\sim3.8$ Gyr$^{-1}$ (M$_\star\sim8\times10^{10}$ M$_\odot$). Gas and dust rich (M$_{\rm gas}\sim4\times10^{10}$ M$_\odot$; M$_{\rm dust}\sim1.4\times10^9$ M$_\odot$), with a typical depletion timescale $\tau_{\rm depl}\sim100$ Myr, its very intense burst of star formation must trigger a rapid growth of the central AGN. 

\subsection{AzTEC.GS21 (J033247.59-274452.3)}

AzTEC.GS21 is a DSFG galaxy at z$_{\rm spec}= 1.910\pm0.001$ \citep[grism redshift by][measured with the FORS2 spectrograph (ESO/VLT), ESO/GOODS spectroscopic campaign program in the GOODS-S field]{Vanzella2008:2008A&A...478...83V}. It was firstly detected at 1.1 mm with AzTEC/ASTE (S$_{\rm 1.1\,mm}=2.7^{+0.6}_{-0.7}$ $\mu$Jy). This photometric data was not included in the SED analysis by \citet[][]{Pantoni2021:2021MNRAS.504..928P} since this flux is probably contaminated by other (sub-)millimeter sources in the neighborhood. Indeed, the H$_{160}$ image of its optical counterpart is disturbed (Fig. \ref{fig:cont}) and \citet{Targett2013:2013MNRAS.432.2012T} claim that AzTEC.GS21 is the primary dominant component of a multiple system. The authors measure an effective H$_{160}$ radius of r$_{\rm H}\sim$ 2.6 kpc and a S\'{e}rsic index n$_H\simeq$ 1.3 (close to a disk-like profile). A more recent analysis on the HST image was performed by \citet{vanderWel2012:2012ApJS..203...24V}, who find a circularized H$_{160}$ radius r$_{\rm H}\sim$ 3.7 kpc. In this work we have exploited the latter result.

The source was detected by ALMA in B7 (at $\lambda=850\,\mu$m) by \citet{Cowie2018:2018ApJ...865..106C}, measuring a flux S$_{850\,\mu m}=3.6\pm0.3$ mJy (source ID: no. 20; name ALMA033247-274452). \citet{Hatsukade2018:2018PASJ...70..105H} measured the source flux in the ALMA B6, reporting the value S$_{\rm B6}=1.86\pm0.32$ mJy.
We found a $>5\sigma$ continuum detection for this source in the ALMA B9 ($\lambda=400-500$ $\mu$m), measuring a flux S$_{\rm    B9}=12\pm1$ mJy (Tab. \ref{tab:continuum}). AzTEC.GS21 is not resolved in the ALMA maps. For this reason we chose the map with the best spatial resolution available in the ALMA Archive in order to give at least an upper limit on its size, i.e. r$_{\rm ALMA}<0.7$ kpc, that we use to trace the bulk of dusty star formation occurring in the galaxy.

The AzTEC source has a radio (unresolved) counterpart (GS21a) in the deep ($\sigma\sim8$ $\mu$Jy) VLA map at 1.4 GHz by \citet{Kellermann2008:2008ApJS..179...71K} and \citet{Miller2013:2013ApJS..205...13M}. The corresponding radio flux is S$_{\rm 1.4\,GHz}=43.6\pm6.3$ $\mu$Jy, that we found to be consistent with the radio emission coming from the host galaxy star formation \citep[][]{Pantoni2021:2021MNRAS.504..928P}.

The AzTEC source has a X-ray counterpart (IDX 852) in the catalog by \citet{Luo2017:2017ApJS..228....2L}. The intrinsic absorption column density $N_H \sim 2.27\times10^{22}$ cm$^{-2}$ gives an absorption-corrected intrinsic $2-10$ keV luminosity of $1.7\times10^{42}$ erg s$^{-1}$. \citet{Luo2017:2017ApJS..228....2L} classified the source as \textit{AGN} (Tab. \ref{tab:galaxy_properties}, even if its X-ray luminosity do not clearly emerge from the host galaxy one \citep[][]{Pantoni2021:2021MNRAS.504..928P}.

The SED analysis by \citet[][]{Pantoni2021:2021MNRAS.504..928P} presents AzTEC.GS21 as a main-sequence galaxy. The galaxy is forming stars at a rate SFR $\sim350$ M$_\odot$ yr$^{-1}$, with a sSFR $\sim2$ Gyr$^{-1}$. Massive (M$_\star\sim2\times10^{11}$ M$_\odot$), gas and dust rich (M$_{\rm gas}\sim5\times10^{10}$ M$_\odot$; M$_{\rm dust}\sim6\times10^8$ M$_\odot$), AzTEC.GS21 is characterized by a depletion timescale of $\tau_{\rm depl}\sim140$ Myr. Notice that its star formation could be fuelled (also) by interactions with some objects in the vicinity, since some works indicate it to be the central, dominant component of a multiple system \citep[][]{Targett2013:2013MNRAS.432.2012T}. However, the clumpy/disturbed morphology of the source in the optical band may also trace the condensation process of the gas phase toward the centre of the galaxy where the dusty star formation is occurring on a radius $<0.7$ kpc. Even accounting the former scenario, we do not expect eventual interactions to have an important impact on the galaxy subsequent evolution, except for prolonging the star-formation, since AzTEC.GS21 should dominate by far the gravitational potential of the system \citep[][]{Targett2013:2013MNRAS.432.2012T}.
We expect the galaxy to be quenched by the central AGN in some hundreds of Myr and subsequently become a red and dead galaxy. Clearly, to confirm the right scenario for the evolution of AzTEC.GS21 we need deeper multi-wavelength images and at higher resolution. 

\subsection{AzTEC.GS22 (J033212.55-274306.1)}

AzTEC.GS22 is a DSFG galaxy firstly detected at 1.1 mm with AzTEC/ASTE (S$_{\rm 1.1\,mm}=2.1\pm0.6$ mJy).
\citet{Yun2012:2012MNRAS.420..957Y} found a red IRAC/MIPS counterpart for the AzTEC source, with a spectroscopic redshift of z$_{\rm spec}= 1.794\pm0.005$ \citep[Tab. \ref{tab:galaxy_properties}][]{Wuyts2009:2009ApJ...706..885W, Targett2013:2013MNRAS.432.2012T}, that we adopt in this work.

We found a $>5\sigma$ continuum detection for AzTEC.GS22 in the ALMA B9 ($\lambda=400-500$ $\mu$m) and we measured a flux S$_{0.45\,mm}=5.8\pm0.8$ mJy (Tab. \ref{tab:continuum}), that was essential to constrain the dusty peak of galaxy SED in \citet[][]{Pantoni2021:2021MNRAS.504..928P}, since the lack of the Herschel 500 $\mu$m photometric point. The ALMA source is not resolved in the continuum B9 thus we provide just an upper limit on its size, i.e. r$_{\rm ALMA}<1.7$ kpc, tracing the bulk of dusty star formation.

The most recent analysis of the optical emission of AzTEC.GS22 was performed by \citet{vanderWel2012:2012ApJS..203...24V}, who measured an optical size of r$_{\rm H}\sim3.2$ kpc, that is more extended than the dust continuum emission (at least of a factor 2).

The faint radio source GS22a, located 7.8 arcsec away from the AzTEC centroid, is adopted in literature as the most likely counterpart at 1.4 GHz \citep{Yun2012:2012MNRAS.420..957Y, Dunlop2017:2017MNRAS.466..861D}. Moreover in \citet[][]{Pantoni2021:2021MNRAS.504..928P} we found that the VLA flux at 1.4 GHz (S$_{\rm 1.4\,GHz}$ = 34.6 $\pm$ 6.5 $\mu$Jy) can be traced back solely to galaxy star formation.

From the SED analysis of AzTEC.GS22 (cf. Tab. \ref{tab:summary_integral}), we obtain that the object is a  main sequence galaxy. The burst of ongoing star formation in this object is almost 1 Gyr old (age $\sim 900-1000$ Myr) and it is forming stars at a rate $\sim90-100$ M$_\odot$ yr$^{-1}$.
Massive (M$_\star\sim6\times10^{10}$ M$_\odot$), gas and dust rich (M$_{\rm gas}\sim5\times10^{10}$ M$_\odot$; M$_{\rm dust}\sim1.4\times10^9$ M$_\odot$), AzTEC.GS22 is characterized by a depletion timescale of $\tau_{\rm depl}\sim550$ Myr.
The huge content of gas and dust could be ascribed  simply to its past star formation activity and/or to some kind of interactions (e.g., gas stripping/harassment) with possible companions observed in the optical by \citet{Targett2013:2013MNRAS.432.2012T}. We do not expect these interactions to have a huge impact on the galaxy, since it must dominate the potential well of the multiple system. To confirm this scenario we need high-resolution multi-wavelength imaging. Indeed, the clumpy/disturbed optical morphology, extending over a radius of a few kpcs, could also be interpreted as a signature of the intense dusty star formation of the galaxy.

In the literature we did not find any X-ray counterpart for the millimeter source, neither in the deep X-ray catalogue by \citet{Luo2017:2017ApJS..228....2L}. Since the source lies in a very deep region of the \textit{Chandra} map (equivalent exposure times of $\gtrsim5$ Ms), we think that the most probable hypothesis to explain the non-detection is that the source is totally obscured in the X-ray. 
Again, to confirm this interpretation we need spatially-resolved images of the system, at least in the radio and millimeter bands, together with a deeper follow-up in the X-ray. 



\bsp	
\label{lastpage}
\end{document}